\documentclass[prl, twocolumn]{revtex4}
\usepackage{graphicx,color}
\usepackage{amssymb}
\usepackage{amsmath}
\usepackage{bm}
\usepackage{hyperref}
\usepackage{tabularx}
\usepackage{multirow}
\usepackage{braket}
\usepackage{xcolor}
\usepackage{gensymb}

\begin{document}

\title{Quantum Oscillation of Thermally Activated Conductivity in a Monolayer WTe$_2$-like Excitonic Insulator}
\author{Wen-Yu He}
\affiliation{Department of Physics, Massachusetts Institute of Technology, Cambridge, Massachusetts 02139, USA}
\author{Patrick A. Lee} \thanks{palee@mit.edu}
\affiliation{Department of Physics, Massachusetts Institute of Technology, Cambridge, Massachusetts 02139, USA}

\date{\today}

\begin{abstract}
Recently, quantum oscillation of the resistance in  insulating monolayer WTe$_2$ was reported. An explanation in terms of  gap modulation in the hybridized Landau levels of an excitonic insulator was also proposed by one of us. However, the previous picture of gap modulation in the Landau levels spectrum was built on a pair of well nested  electron and hole Fermi surfaces, while  the monolayer WTe$_2$  has one hole and two electron Fermi pockets with relative anisotropy.  Here we demonstrate that for system like monolayer WTe$_2$, the excitonic insulating state arising from the coupled one hole and two electron pockets possesses a finite region in interaction parameter space that shows gap modulation in a magnetic field. In this region, the thermally activated conductivity displays the $1/B$ periodic oscillation and it can further develop into discrete peaks at low temperature,  in  agreement with the experimental observation. We show that the relative anisotropy of the bands is a key parameter and the quantum oscillations decrease rapidly if the anisotropy increases further than the realistic value for monolayer WTe$_2$.
\end{abstract}

\maketitle

\emph{Introduction}. Monolayer WTe$_2$ is a  two-dimensional crystal that exhibits many unusual properties: it is a quantum spin Hall insulator up to relatively high temperature and is a superconductor when doped ~\cite{Xiaofeng, Tang, Cobden, Sanfeng1, Fatemi, Sajadi}. Recently, quantum oscillation (QO) of the resistance was reported in monolayer WTe$_2$ when it is gate tuned to be an insulator~\cite{Sanfeng2}. In a separate paper, it was argured that the insulating state of monolayer WTe$_2$ is an excitonic insulator~\cite{Sanfeng3}. These observations in monolayer WTe$_2$ put it into the category of the insulators like SmB$_6$~\cite{Sebastian} and YbB$_{12}$~\cite{LuLi}, which possess QOs along with insulating electrical conductivity. As the canonical understanding of QOs is based on the existence of electron Fermi surface (FS) in materials~\cite{Shoenberg}, the origin of QO observed in the insulating monolayer WTe$_2$ requires an explanation.

\begin{figure}
\centering
\includegraphics[width=3.5in]{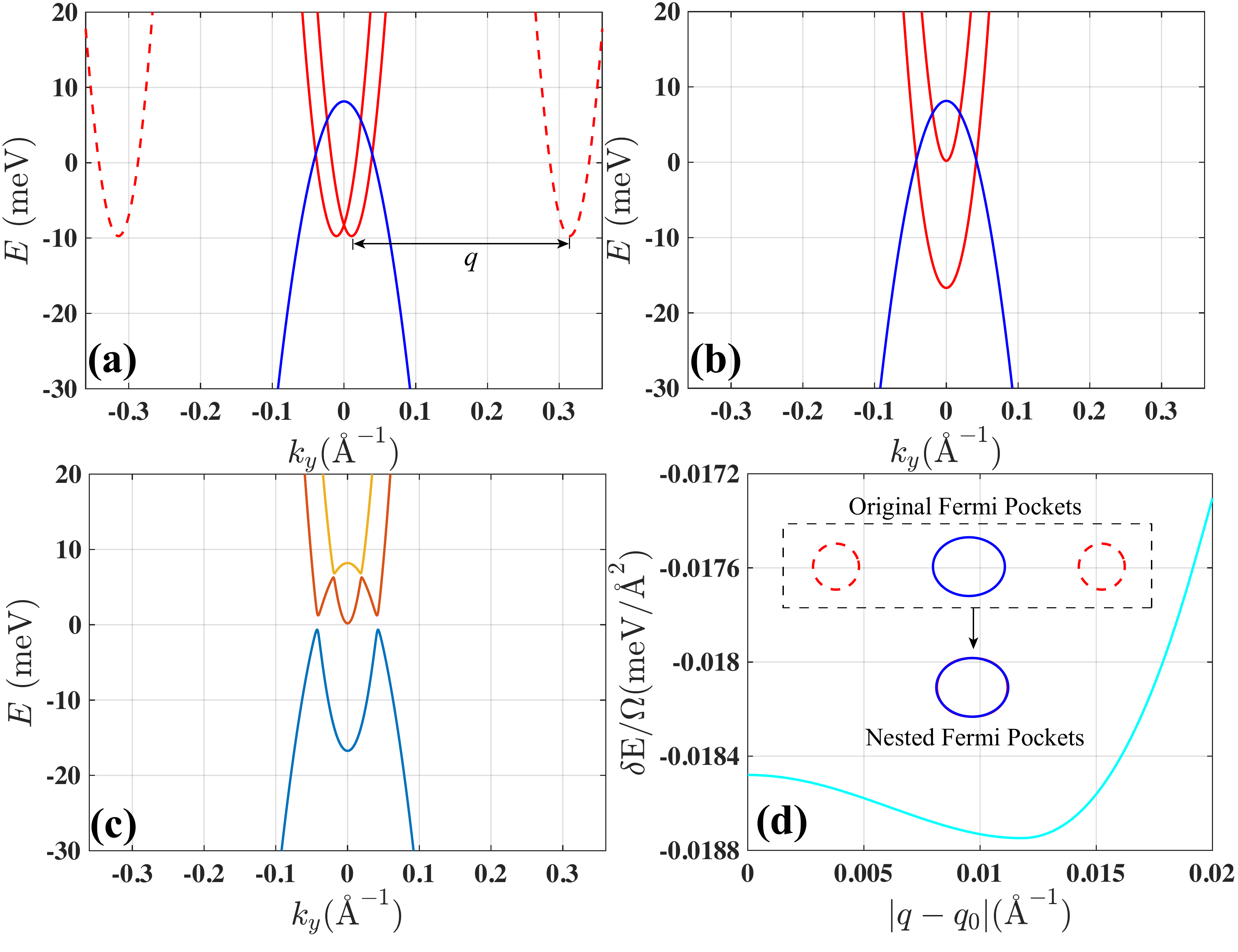}
\caption{(a) Band structure of monolayer WTe$_2$. The red dashed lines denote the original position of conduction bands. Due to the density modulation in the excitonic insulating state, the original conduction bands are shifted to $\bm{k}=\bm{0}$ by $\pm\bm{q}$. (b) The energy dispersions when the two conduction bands in (a) have finite hybridization $V_2$ but no coupling $\left(V_1=0\right)$ with the valence band. They are split into upper and lower bands. (c) The insulating gap generated from (b) by setting the coupling $V_1\neq 0$. (d) The energy density gained in the insulating state as a function of the deviation of $q$ from $q_0$. Here $\Omega$ is the sample area and the  energy gained is $\delta E=\left\langle E \right\rangle -\left\langle E \right\rangle_0$ with $\left\langle E \right\rangle_0$ being the ground state energy at $\left(V_1, V_2\right)=\left(0, 0\right)$ meV. The inset inside the dashed rectangle shows the original electron and hole Fermi pockets. At the optimal shift momentum $\bm{q}$ with $V_2=-7.9$ meV, the electronic pocket from lower band in (b) gets well nested with the hole pocket. In the calculations the chemical potentials are taken to be $\mu_{\textrm{c}}=9.7$ meV and $\mu_{\textrm{v}}=8.1$ meV.}\label{figure1}
\end{figure} 

The paradox of QO in the  insulating monolayer WTe$_2$ without electron FS has inspired suggestions of neutral FS existing inside the insulating gap, where the neutral FS is formed from itinerant spinons that result from the spin-charge separation of holes~\cite{Sanfeng2}. The neutral FS hypothesis requires the electronic correlation in monolayer WTe$_2$ to be strong enough to fractionalize the holes into spinons and holons~\cite{Motrunich, Senthil1, Senthil2}. On the other hand, a more conventional but still nontrivial explanation that does not appeal to strong electronic correlation in monolayer WTe$_2$ was proposed by one of us: the observed QO of resistance can arise from a small modulation of the insulating gap induced by magnetic field~\cite{Lee1}. In the phenomenological model, the gap modulation can give rise to oscillation of thermally activated conductivity and the oscillation further evolves into sharp periodic spikes in low temperature~\cite{Lee1}, which matches qualitatively well with the temperature dependent behavior of conductance oscillation in monolayer WTe$_2$. In monolayer WTe$_2$ the bandwidth is not particularly narrow and  strong electronic correlation is not expected in this material; so the  modulation of excitonic insulating gap by magnetic field is an attractive alternative explanation for the QO in the insulating monolayer WTe$_2$. Since the origin of QO reported in WTe$_2$ is still under debate, we will proceed with further analysis of the gap modulation model, which is an interesting theoretical problem in its own right.

The scenario of magnetic field modulated insulating gap was first proposed to explain the QO of magnetization observed in the Kondo insulator SmB$_6$~\cite{Cooper1, FaWang, Cooper2}, and later it successfully predicted the QO of conductance in the inverted narrow gap regime of InAs/GaSb quantum wells~\cite{Cooper3, Samarth, RuiRuiDu}. In both SmB$_6$ and InAs/GaSb quantum wells, the insulating gap arises from the hybridization of a single electron band which overlaps a single hole band. Since both are assumed to be isotropic, the system is perfectly nested and fully gapped by the hybridization. Prior to hybridization, the Landau levels in the presence of a magnetic field $B$ moves up in energy in the electron band and down in the hole band. These collide in a periodic way as a function of $1/B$. The memory of this periodicity is retained in the energy gap after hybridization, leading to a periodic modulation of the hybridization gap~\cite{Cooper1, FaWang, Cooper2, Cooper3}. However, in the case of monolayer WTe$_2$, there are one hole and two electron Fermi pockets with relative anisotropy~\cite{Xiaofeng, Sanfeng3, Neupert, JustinSong, Sun}, and in general these pockets are not well nested. The distinctive type of band structure of monolayer WTe$_2$ shown in Fig. \ref{figure1} complicates the application of the gap modulation scenario, raising the question of whether QO oscillations survive in the case of the coupled three Fermi pockets in monolayer WTe$_2$.

In this work, we show that for a monolayer WTe$_2$-like excitonic insulator, there exists a reasonable range of parameter space which exhibits QO of thermally activated conductivity that originates from the magnetic field induced gap modulation.  In the excitonic insulating state, the two electron pockets are first shifted to the middle hole pocket and the mutual couplings among the three Fermi pockets gap out the FSs. Suppose $V_1$ is the coupling between the electron and hole pockets and $V_2$ is the hybridization between the two electronic pockets, the system will self-tune to  the optimal shift momentum $\pm\bm{q}$ that minimizes the ground state energy. Note that in general  $\bm{q}$ is different from $\bm{q_0}$ which is the separation between the top of the hole band and the bottom of the conduction band. We first minimize the energy with respect to the shift vector $|\bm{q}|$, and then construct the phase diagram in the parameter space of $\left(V_1, V_2\right)$. We find that the insulating regime has the ``$<$" shape looking, and near the tip area of ``$<$" the insulating gap has reasonable modulation (see Fig. \ref{figure2} and \ref{figure3}). We find that the area in the phase diagram that can generate finite gap modulation depends on the relative anisotropy of Fermi pockets: it shrinks as the relative anisotropy increases. For the monolayer WTe$_2$ with the relative anisotropy estimated to be around 1.5~\cite{Sun}, we find that the gap modulation near the tip area of ``$<$" can indeed give rise to observable conductivity oscillation (see Fig. \ref{figure3} and \ref{figure4}), so it gives the possibility that the gap modulation scenario can explain the QO in monolayer WTe$_2$.

\emph{Model for the excitonic insulating state}. The monolayer WTe$_2$ has a hole Fermi pocket centered at $\bm{k}=\bm{0}$ and two flanking electronic Fermi pockets at $\bm{k}=\pm\bm{q}_0$ with $\bm{q}_0=\left(0, q_0\right)$, as is shown in the inset of Fig. \ref{figure1} (d). In the band basis, the quadratic dispersions are taken to approximate the conduction band as  $\epsilon_{\pm}\left(\bm{k}\right)\approx\frac{\hbar^2k_x^2}{2m_{\textrm{c}, x}}+\frac{\hbar^2\left(k_y\mp q_0\right)^2}{2m_{\textrm{c}, y}}-\mu_{\textrm{c}}$, and also the valence band as $\epsilon_{\textrm{v}}\left(\bm{k}\right)\approx-\frac{\hbar^2k_x^2}{2m_{\textrm{v}, x}}-\frac{\hbar^2k_y^2}{2m_{\textrm{v}, y}}+\mu_{\textrm{v}}$. The band masses are fit to be $m_{\textrm{c}, x}=m_{\textrm{c}, y}=0.29m_{\textrm{e}}$, $m_{\textrm{v}, x}=\frac{2}{3}m_{\textrm{v}, y}=0.56m_{\textrm{e}}$~\cite{Sun}. Here the relative anisotropy of the electron and hole pockets is controlled by $\gamma=\frac{m_{\textrm{v}, y}m_{\textrm{c}, x}}{m_{\textrm{v}, x}m_{\textrm{c}, y}}$. Note that since it is always possible to rescale the $k_y$ axis so that one of the the bands is isotropic, even in the presence of a magnetic field, it is only the relatively anisotropy that matters in the discussion that follows.    When the Coulomb interaction effect is neglected, the conduction band stays at its original position as is indicated by the red dashed line in Fig. \ref{figure1} (a).

The role of Coulomb interaction in the system with energy overlap between the conduction and valence bands is to provide an effective inter-band attraction to bind electrons and hole states into excitons~\cite{Kohn1, Kohn2, Rice1, Rice2}. In an excitonic insulator,  a density modulation at $\bm{q}$ is spontaneously generated, so the original electron and hole Fermi pockets are shifted  to  gap out the entire FS. For the monolayer WTe$_2$ type of band structure which has two conduction band minima and one valence band maximum, besides the coupling between the conduction and valence bands, the hybridization between the two electron pockets is also needed to generate the insulating gap~\cite{Lee1}. As a result, the generic mean field Hamiltonian matrix for the excitonic insulating state is assumed to have the form
\begin{align}\label{bare_Hamiltonian}
H=\begin{pmatrix}
\epsilon_{\textrm{v}}\left(\bm{k}\right) & V_1 & V_1 \\
V_1 & \epsilon_{+}\left(\bm{k}+\bm{q}\right) & V_2 \\
V_1 & V_2 & \epsilon_{-}\left(\bm{k}-\bm{q}\right)
\end{pmatrix},
\end{align}
where the basis is $\left[\psi_{\textrm{v}, \bm{k}}, \psi_{+, \bm{k}+\bm{q}}, \psi_{-, \bm{k}-\bm{q}}\right]^{\textrm{T}}$ with $\psi_{\textrm{v}, \bm{k}}$, $\psi_{\pm, \bm{k}}$ to annihilate a state at $\bm{k}$ in the valence and conduction bands respectively. Here the coupling potential $V_1$ is from the pseudo-spin density order at $\bm{q}$ while the hybridization $V_2$ comes from the charge density order at $2\bm{q}$, which can be obtained from the Hartree-Fock mean field calculation~\cite{Parameswaran}. The pseudo-spin index is dropped as it does not affect the energy eigenvalue~\cite{Supp}. Due to the spontaneous generated density order, the original conduction bands are first shifted to $\bm{k}=\bm{0}$ as is seen in Fig. \ref{figure1} (a). Then the potential $V_2$ hybridizes the two conduction bands, generating two new conduction bands with different energies shown in Fig. \ref{figure1} (b). We refer to these as the upper and lower bands. Finally, the valence band couples with the two new conduction bands individually and an excitonic insulating gap appears in Fig. \ref{figure1} (c). Importantly, for negative $V_2$ the coupling between the valence band and the lower conduction band  is dominant in the insulating gap generation~\cite{Supp} while for positive $V_2$ it is the opposite. Clearly negative $V_2$ is preferable for energy gain and our calculations are mainly in this domain. Since a microscopic  model for the interaction parameters is not known, in this paper we do not attempt to perform a self-consistent calculation starting with a microscopic model, but consider only the parameter space controlled by the mean field parameters $V_1$ and $V_2$.

For a given $V_1$ and $V_2$ , the system will self-tune to have the ground state energy $\left\langle E \right\rangle=\sum_{\bm{k}, \nu}f\left[E_{\nu}\left(\bm{k}\right)\right]E_{\nu}\left(\bm{k}\right)$ minimized by varying the shift momentum $\bm{q}=\left(0, q\right)$. Here $f\left(\epsilon\right)$ is the Fermi Dirac distribution function and $E_{\nu}\left(\bm{k}\right)$ with $\nu=1, 2, 3$ is the energy eigenvalue of the Hamiltonian $H$ in Eq. \ref{bare_Hamiltonian}. In the simple case when there is no relative anisotropy $\left(\gamma=1\right)$, for sufficient large $V_2$ the ground state energy gets minimized at $\bm{q}=\bm{q}_0$ when only the  electron pocket corresponding to the lower band crosses the Fermi level and is perfectly nested with the hole pocket~\cite{Supp}. In the realistic monolayer WTe$_2$ band structure, the middle hole pocket is more elliptical than the flanking electron pockets, so it is no longer true that the optimal nesting always occurs at $\bm{q}=\bm{q}_0$. Instead the $\bm{q}$ may be adjusted so that one side of the electron pocket has optimal overlap with a side of the hole pocket. In Fig. \ref{figure1} (d), the energy density gained in the insulating state is plotted as a function of $|\bm{q}-\bm{q}_0|$ with $\left(V_1, V_2\right)=\left(1.3, -7.9\right)$ meV, which shows that the ground state energy takes the minimal value at nonzero $|\bm{q}-\bm{q}_0|$. By calculating the insulating gap size and the associated shift momentum $\bm{q}$ in a range of the coupling potentials $V_1$, $V_2$, the excitonic insulating phase diagram in terms of $\left(V_1, V_2\right)$ is  obtained in Fig. \ref{figure2} using realistic parameters for monolayer WTe$_2$. In the phase diagram, the gapped regime has a pointy ``$<$" shape, and the finite $|\bm{q}-\bm{q}_0|$ near the tip of ``$<$" makes the $V_2$ hybridized electronic pockets well nested with the middle hole pocket as can be seen in the inset of Fig. \ref{figure1} (d). As we shall see, it is in this well nested region that the gap modulation and QO in the presence of a magnetic field survives the best.

\begin{figure}
\centering
\includegraphics[width=3.6in]{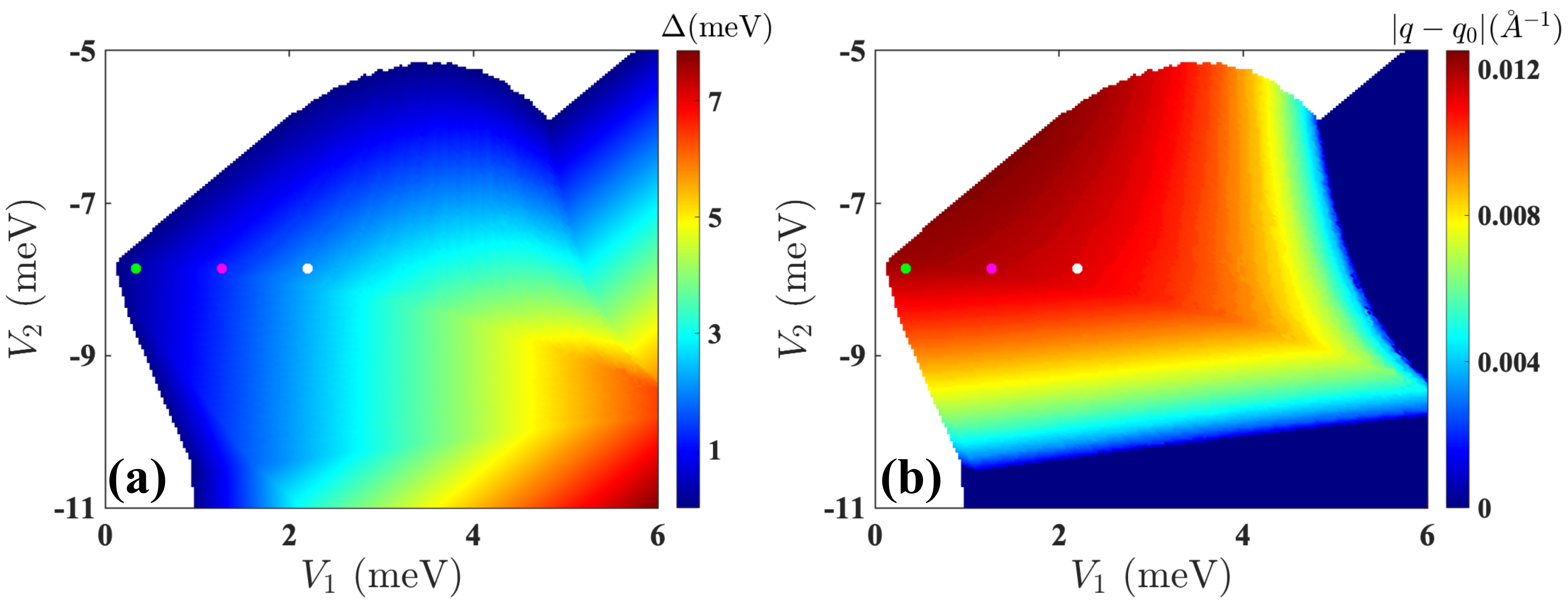}
\caption{The insulating phase diagram of a monolayer WTe$_2$-like excitonic insulator using realistic band parameters which sets the relative anisotropy to be  $\gamma = 1.5$. At given coupling potential $\left(V_1, V_2\right)$, the gap size $\Delta$ is given in (a) and the associated shift momentum $|q-q_0|$ is present in (b). }\label{figure2}
\end{figure}

\emph{Gap modulation in the Landau levels spectrum}. In the presence of magnetic field, the original quadratic energy dispersions for the conduction and valence bands form three sets of Landau levels: $\epsilon_{\pm, n}=\left(n+\frac{1}{2}\right)\hbar\omega_{\textrm{c}}-\mu_{\textrm{c}}$, $\epsilon_{\textrm{v}, n}=-\left(n+\frac{1}{2}\right)\hbar\omega_{\textrm{v}}+\mu_{\textrm{v}}$, with $n\in\mathbb{Z}$.  Here the cyclotron frequencies are $\omega_{\textrm{c}}=\frac{eB}{m_{\textrm{c}}}$, $\omega_{\textrm{v}}=\frac{eB}{m_{\textrm{v}}}$, with the cyclotron masses being $m_{\textrm{c}}=\sqrt{m_{\textrm{c}, x}m_{\textrm{c}, y}}$, $m_{\textrm{v}}=\sqrt{m_{\textrm{v}, x}m_{\textrm{v}, y}}$. In the excitonic insulating state, due to the couplings among the original conduction and valence bands, the formed Landau levels get hybridized, resulting in an insulating gap as a function of magnetic field $B$. The Hamiltonian matrix that involves Landau levels hybridizations has similar form as that in Eq. \ref{bare_Hamiltonian}~\cite{Supp}
\begin{align}\label{Hamiltonian_LL}
\hat{H}=&\begin{pmatrix}
\hat{H}_{\textrm{v}} & \hat{V}_1 & \hat{V}_1\\
\hat{V}^\dagger_1 & \hat{H}_+ & \hat{V}_2 \\
\hat{V}^\dagger_1 & \hat{V}^\dagger_2 & \hat{H}_-
\end{pmatrix},
\end{align}
where the matrix elements are given in Landau gauge $\bm{A}=\left(0, Bx, 0\right)$ as
\begin{widetext}
\begin{align}
\hat{H}_{\textrm{v}, n, m}=&\bra{\textrm{v}, n}-\frac{1}{2}\hbar\omega_{\textrm{v}}\left(a^\dagger_{\textrm{v}}a_{\textrm{v}}+a_{\textrm{v}}a^\dagger_{\textrm{v}}\right)+\mu_{\textrm{v}}\ket{\textrm{v}, m},\\\nonumber
\hat{H}_{\pm, n, m}=&\bra{\textrm{c}, n}\frac{1}{2}\hbar\omega_{\textrm{c}}\left\{\left[a^\dagger_{\textrm{c}}\mp\frac{l_{\textrm{c}, B}}{\sqrt{2}}\left(q-q_0\right)\right]\left[a_{\textrm{c}}\mp\frac{l_{\textrm{c}, B}}{\sqrt{2}}\left(q-q_0\right)\right]\right.\\
&\left.+\left[a_{\textrm{c}}\mp\frac{l_{\textrm{c}, B}}{\sqrt{2}}\left(q-q_0\right)\right]\left[a^\dagger_{\textrm{c}}\mp\frac{l_{\textrm{c}, B}}{\sqrt{2}}\left(q-q_0\right)\right]\right\}-\mu_{\textrm{c}}\ket{\textrm{c}, m},\\
\hat{V}_{1, n, m}=&\bra{\textrm{v}, n}V_1\ket{\textrm{c}, m},\quad \textrm{and}\quad\hat{V}_{2, n, m}=\bra{\textrm{c}, n}V_2\ket{\textrm{c}, m},
\end{align}
\end{widetext}
with $l_{\textrm{c}, B}=\sqrt{\frac{\hbar}{m_{\textrm{c}, x}\omega_{\textrm{c}}}}$. Here the basis $\ket{\textrm{v}, n}=\frac{\left(a^\dagger_{\textrm{v}}\right)^n}{\sqrt{n!}}\ket{\textrm{v}, 0}$ and $\ket{\textrm{c}, n}=\frac{\left(a^\dagger_{\textrm{c}}\right)^n}{\sqrt{n!}}\ket{\textrm{c}, 0}$ are the eigenstates corresponding to the $n$th Landau level from the valence band $\epsilon_{\textrm{v}}\left(\bm{k}\right)$ and the $\bm{k}=\bm{0}$ centered conduction band $\epsilon_{\textrm{c}}\left(\bm{k}\right)=\frac{\hbar^2k_x^2}{2m_{\textrm{c}, x}}+\frac{\hbar^2k_y^2}{2m_{\textrm{c}, y}}-\mu_{\textrm{c}}$ respectively. It is clear that $\hat{H}_{\textrm{v}}$ and $\hat{H}_\pm$ gives energy eigenvalues $\epsilon_{\textrm{v}, n}$ and $\epsilon_{\pm, n}$ respectively. Suppose the FS area covered by the electron pockets and hole pocket are both equivalent to $S$, which is required by charge neutrality, then the chemical potentials for the original conduction and valence bands are solved to be $\mu_{\textrm{c}}=\frac{\hbar^2S}{4\pi m_{\textrm{c}}}$, $\mu_{\textrm{v}}=\frac{\hbar^2S}{2\pi m_{\textrm{v}}}$. It indicates that once the cyclotron masses $m_{\textrm{c}}$, $m_{\textrm{v}}$ and the FS area $S$ are fixed, the energy eigenvalue of the Hamiltonian in Eq. \ref{Hamiltonian_LL} depends only on the specific forms of the hybridization matrix $\hat{V}_1$, $\hat{V}_2$. In the case that Fermi pockets have no relative anisotropy $\left(\gamma=1\right)$, the hybridization matrix $\hat{V}_1$, $\hat{V}_2$ are both diagonal, so the energy eigenvalue are analytically solved to be~\cite{Supp}
\begin{align}
E_{1, n}=&\frac{1}{2}\left(\tilde{\epsilon}_{\textrm{c}, n}+\epsilon_{\textrm{v}, n}\right)+\sqrt{\frac{1}{4}\left(\tilde{\epsilon}_{\textrm{c}, n}-\epsilon_{\textrm{v}, n}\right)^2+2V_1^2},\\
E_{2, n}=&\frac{1}{2}\left(\tilde{\epsilon}_{\textrm{c}, n}+\epsilon_{\textrm{v}, n}\right)-\sqrt{\frac{1}{4}\left(\tilde{\epsilon}_{\textrm{c}, n}-\epsilon_{\textrm{v}, n}\right)^2+2V_1^2},\\
E_{3, n}=&\hbar\omega_{\textrm{c}}\left(n+\frac{1}{2}\right)-V_2,
\end{align}
with $\tilde{\epsilon}_{\textrm{c}, n}=\hbar\omega_{\textrm{c}}\left(n+\frac{1}{2}\right)+V_2$. The resulting Landau levels spectrum always have the magnetic field modulated gap $\sqrt{\left(\tilde{\epsilon}_{\textrm{c}, n}-\epsilon_{\textrm{v}, n}\right)^2+8V_1^2}$ and the modulation periodicity is determined by the FS area before hybridization~\cite{Cooper1, Supp}. In the more general case of monolayer WTe$_2$ type band structure that has relative anisotropy $\gamma>1$, the off-diagonal elements in the hybridization matrix $\hat{V}_1$ are generally nonzero (the detailed calculations for the matrix elements of $\hat{V}_1$ are present in the Supplemental Material~\cite{Supp}). As numerically diagonalizing $\hat{H}$ gives the the hybridized Landau levels spectrum in the case of $\gamma>1$, the effect of relative anisotropy on the magnetic field induced gap modulation can be figured out.

\begin{figure}
\centering
\includegraphics[width=3.6in]{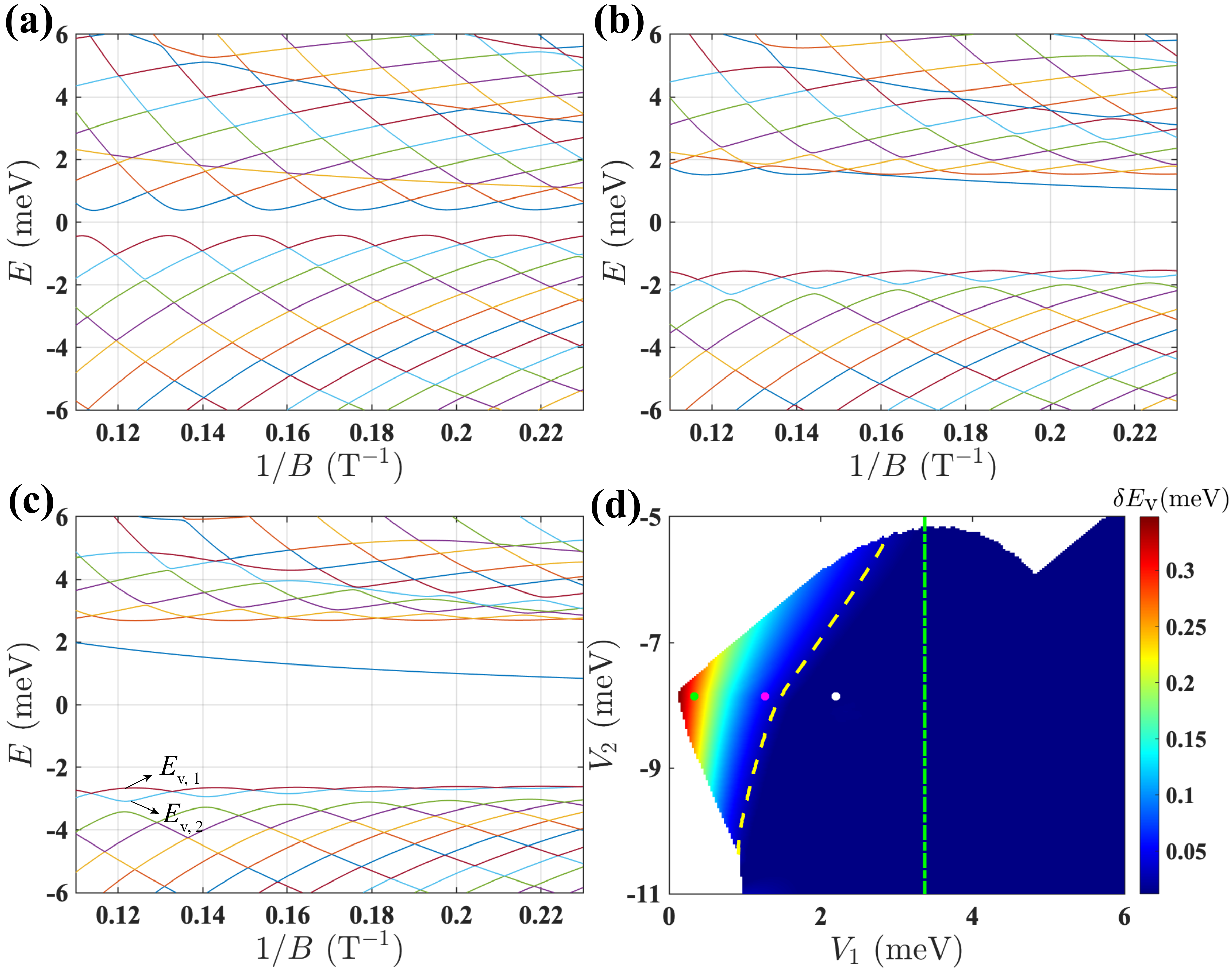}
\caption{The Landau levels spectrum from the green, magenta and white phase points in Fig. \ref{figure2} for (a), (b) and (c) respectively. Notice that a band which corresponds to the upper electron band in Fig. \ref{figure1} (b) is very weakly hybridized and stays at the same energy. It moves inside the gap when the gap opens for increasing $V_1$. It shows no modulation with $B$. On the other hand, the top of the valence band always shows modulations, even though it is weakened for increasing $V_1$. (d) The gap modulation of the valence band is characterized by the energy difference $\delta E_{\textrm{v}}$ at given $\left(V_1, V_2\right)$ in the gapped region. The yellow dashed line is the contour of $\delta E_{\textrm{v}}=0.03$ meV. Recall that the relative anisotropy is $\gamma = 1.5$. For comparison, the green dash-dot line gives the boundary of the region with gap modulation larger than 0.03 meV in $\gamma=1$ case.}\label{figure3}
\end{figure}

For the monolayer WTe$_2$, the electron and hole cyclotron masses take the value $m_{\textrm{c}}=0.29 m_{\textrm{e}}$, $m_{\textrm{v}}=0.67 m_{\textrm{e}}$~\cite{Sun}. Given the experimental observed QO frequency $f=48.6$ T in device 1~\cite{Sanfeng2}, the FS area $S$ can be determined by the Onsager theorem $f=\frac{\hbar S}{2\pi e}$, so the chemical potentials are fixed to be $\mu_{\textrm{c}}=9.7$ meV, $\mu_{\textrm{v}}=8.1$ meV. The set of parameters $m_{\textrm{c}}$, $m_{\textrm{e}}$, $\mu_{\textrm{c}}$ and $\mu_{\textrm{v}}$ along with the relative anisotropy $\gamma=1.5$ have been applied in the calculation for the phase diagram present in Fig. \ref{figure2}. Three points with the same $V_2$ but different $V_1$ in the phase diagram are selected to calculate the Landau levels spectrum. The Landau levels spectrum from the green, magenta, and white colored phase points in Fig. \ref{figure2} are plotted as a function of $1/B$ in Fig. \ref{figure3} (a), (b) and (c) respectively. The hybridization of Landau levels described by Eq. \ref{Hamiltonian_LL} inherits the feature of the three bands coupling in Eq. \ref{bare_Hamiltonian}. The hybridization matrix $\hat{V}_2$ first lifts the degeneracy of $\epsilon_{\pm, n}$, giving two sets of Landau levels that are from the two new conduction bands in Fig. \ref{figure1} (b). Then the newly generated two sets of Landau levels couple individually with the valence band Landau levels $\epsilon_{\textrm{v}, n}$, eventually generating the gap in the Landau levels spectrum. In Fig. \ref{figure3} (a), with given $\left(V_1, V_2\right)$ near the tip of the gapped region, the gap in the Landau levels spectrum shows significant modulation in the magnetic field. When the coupling potential $V_1$ increases, the insulating gap becomes larger so the lowest Landau level from the upper conduction band  in Fig. \ref{figure1} (b) appears inside the gap at small $B$. Since that Landau level has negligible coupling with the valence band Landau levels $\epsilon_{\textrm{v}, n}$, the energy oscillation in the upper boundary of the gap disappears there. On the other hand, the energy oscillation in the lower boundary of the gap survives in the whole range of magnetic field $1/B\in\left[0.11,  0.23\right]$ T$^{-1}$ shown in Fig. \ref{figure3} (b), although it is suppressed a bit due to larger $V_1$. As the coupling potential $V_1$ further increases, the energy oscillation in the lower boundary gradually gets smoothed in $1/B>0.22$ T$^{-1}$ as is shown in Fig. \ref{figure3} (c).

The energy difference between the top  two valence band Landau levels at $E_{\textrm{v}, 1}$, $E_{\textrm{v}, 2}$ serves as an indicator of the gap modulation. For the Landau levels spectrum in the range of magnetic field $1/B\in\left[0.11, 0.23\right]$ T$^{-1}$, the energy difference $\delta E_{\textrm{v}}=\frac{1}{2}\left[E_{\textrm{v}, 1}\left(1/B\right)-E_{\textrm{v}, 2}\left(1/B\right)\right]$ at $1/B=0.23$ T$^{-1}$ is calculated in the gapped region, which is shown in Fig. \ref{figure3} (d). Importantly, the energy difference $\delta E_{\textrm{v}}$ is found to decrease from a finite value to zero as $\left(V_1, V_2\right)$ goes away from the tip area of ``$<$", and  the gap modulation decays in the same way. In Fig. \ref{figure3} (d), the contour of $\delta E_{\textrm{v}}=0.03$ meV gives an estimate of the regime that has reasonable gap modulation, and the gap modulation is further confirmed by Landau levels spectrum calculations from more  points inside the regime $\delta E_{\textrm{v}}>0.03$ meV~\cite{Supp}. Compared to the $\gamma=1$ case where the $V_1< 3.4$ meV delimits the area of gap modulation larger than $0.03$ meV~\cite{Supp}, the relative anisotropy $\gamma=1.5$ reduces the area in the phase diagram. Besides the specific case of $\gamma=1.5$, the energy difference $\delta E_{\textrm{v}}$ in the phase diagram has been calculated for a series of insulating states with $\gamma=2, 2.5, 3$ in the Supplemental Material~\cite{Supp}. The area with reasonable gap modulation is found to shrink toward the tip of ``$<$" as the relative anisotropy increases. The reason is that as $\gamma$ increases, the region in parameter space where the electron and hole Fermi surfaces are well nested prior to hybridization by $V_1$ shrinks. When the Fermi surfaces are not well nested, each Landau level in the hole band is coupled to several other ones in the conduction bands when the magnetic field is turned on,  and the gap modulation decreases. Nevertheless,  up to $\gamma=3$, a range of phase space always exists to give gap modulation in the order of $0.1$ meV at $1/B=0.23$ T$^{-1}$, which is enough to generate QO of thermally activity in the range of magnetic field $B\in\left[3, 10\right]$ T~\cite{Supp}.

\begin{figure}
\centering
\includegraphics[width=3.2in]{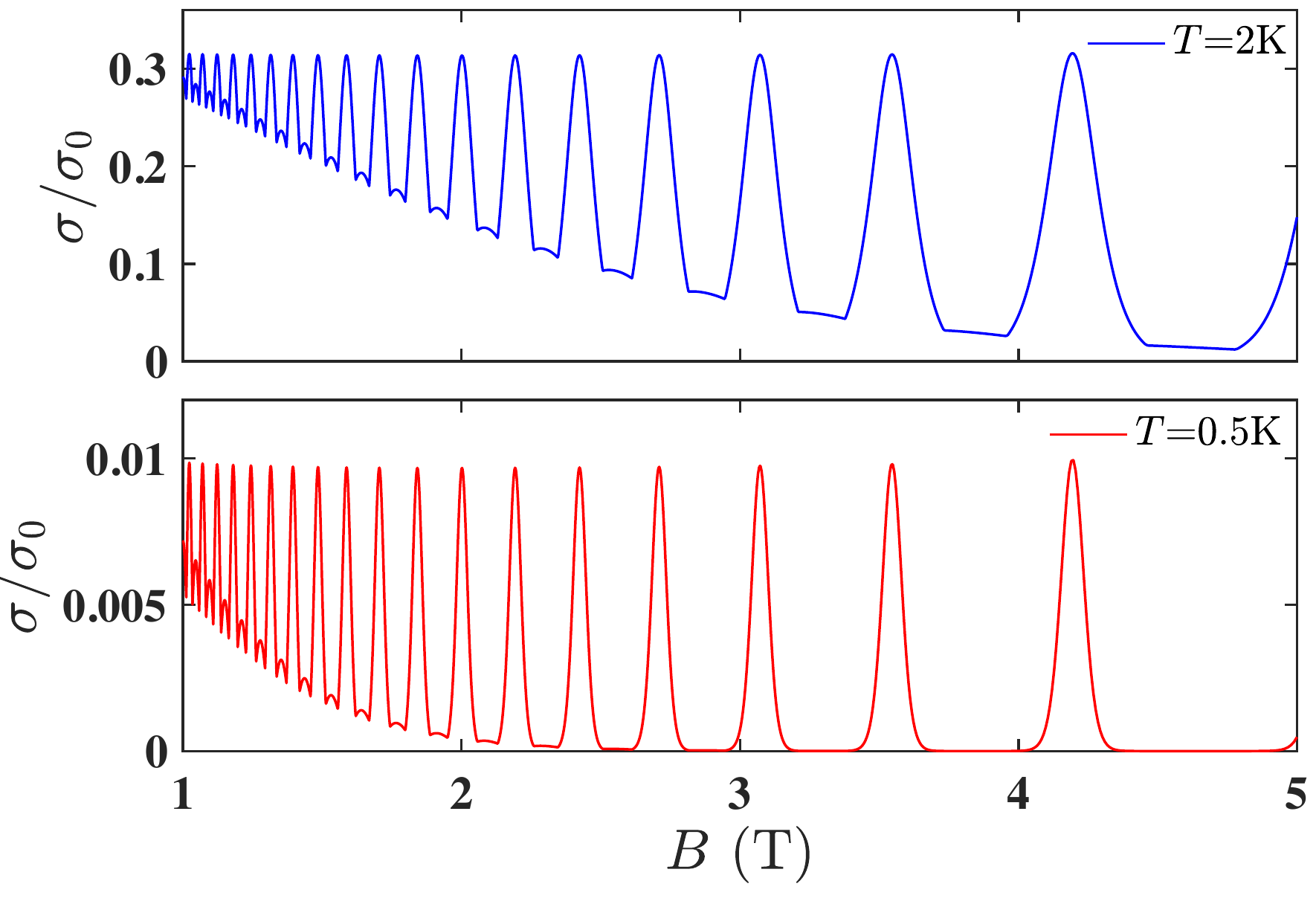}
\caption{The thermally activated conductivity calculated from Eq. \ref{act_conductivity}. In the calculation the chemical potentials are fixed to be $\mu_{\textrm{c}}=4.6$ meV, $\mu_{\textrm{v}}=3.9$ meV so that the Fermi surface area gives the frequency $f=\frac{\hbar S}{2\pi e}=23$ T.}\label{figure4}
\end{figure}

\emph{Thermally activated conductivity oscillation}. In the excitonic insulating state which occurs at the charge neutrality, the chemical potential always stays inside the hybridization gap, so the charge carriers at finite temperature are the thermally activated electronic states that come from the Landau levels below the gap. The activated conductivity is described by the Arrhenius type equation~\cite{Shklovskii1, Shklovskii2}:
\begin{align}\label{act_conductivity}
\sigma=\sigma_0\exp\left[-\frac{\Delta\left(B\right)-\Delta\left(0\right)}{k_{\textrm{b}}T}\right],
\end{align}
where $\Delta\left(B\right)$ is the gap at $B$, and $k_{\textrm{b}}$ is the Boltzmann constant. Here the prefactor $\sigma_0$ is assumed to be independent of $B$ as the chemical potential pinned inside the hybridization gap does not oscillate with the applied magnetic field. In the Landau levels spectrum of a monolayer WTe$_2$-like excitonic insulator, since the gap $\Delta\left(B\right)$ is periodically modulated by the magnetic field $B$ in a range of phase space, the associated thermally activated conductivity also exhibits periodic oscillation in $1/B$. In Fig. \ref{figure4}, the thermally activated conductivity is calculated using the Landau levels spectrum that has periodic gap modulation with the same frequency as observed in the conductance oscillation in device 2 in experiment~\cite{Sanfeng2}. As the ratio of conductance between $T=2$ K and $T=0.5$ K in device 2 is smaller than $10^2$, the gap estimated from Eq. \ref{act_conductivity} has the upper limit at the order of 0.2 meV. Hence we take the coupling potentials in the Landau levels spectrum to be $\left(V_1, V_2\right)=\left(0.1, 3.7\right)$ meV so that the Landau levels hybridization gap is around 0.2 meV. The resulting conductivity in Fig. \ref{figure4} clearly shows the oscillation.  At the lower temperature $T=0.5$ K, the conductivity oscillation evolves into periodic spikes that resembles the discrete peaks observed in the quantized regime in experiment. It matches the fact that lowering temperature makes the quantized regime accessible to the range of magnetic field applied in experiment. In the Supplemental Material~\cite{Supp}, thermally activated conductivity has been considered in a range of parameters and the conductivity oscillation is found to be a general phenomenon that will occur in the phase space with visible gap modulation.

\emph{Discussion and Summary}. In the above sections, the effect of impurities that are always contained in the sample has yet been analyzed, but the scenario of QO from the gap modulation would be the same given the impurity potentials are weak. We know that impurities will lift the degeneracy of each Landau level so that the resulting Landau levels get broadened. Prior to the excitonic hybridization, the broadened Landau levels form peaks in the density of states consisting of extended states, while the states between the peaks are localized. When the excitonic hybridization is turned on, the extended states from the electron band would couple with the extended states in the hole bands to form new extended states.  The localized states will also hybridize but remain localized. The resulting hybridized Landau levels in the excitonic insulating state are therefore also broadened in the same way, with extended and localized states. As long as the broadening is smaller than the Landau level spacing, the conductivity from the thermally excited extended states will oscillate due to the gap modulation.

To summarize, for a monolayer  excitonic  insulator that has one hole and two electronic Fermi pockets similar to that of WTe$_2$, there exists a range of phase space that can generate QO of thermally activated conductivity. The size of the phase space with finite QO depends on the relative anisotropy of the Fermi pockets before hybridization. For a relative anisotropy $\gamma$ estimated up to 3, our study shows that the gap modulation survives near the tip of the gapped region in phase diagram. Since the mean field parameters $V_1$, $V_2$ in the phase diagram depends on details of interactions, we do not know whether a self-consistent calculation starting from the electron interactions in the monolayer WTe$_2$ will land us in that region. Thus while it is possible that the gap modulation scenario can explain the QO observed in monolayer WTe$_2$, it is not guaranteed to be always the case.

\emph{Acknowledgements}. The authors acknowledge the support by DOE office of Basic Sciences grant number DE-FG02-03-ER46076.

\onecolumngrid
\clearpage
\begin{center}

{\bf Supplementary Material for ``Quantum Oscillation of Thermally Activated Conductivity in a Monolayer WTe$_2$-like Excitonic Insulator''}

\end{center}

\maketitle
\setcounter{equation}{0}
\setcounter{figure}{0}
\setcounter{table}{0}
\setcounter{page}{1}
\makeatletter

\renewcommand{\theequation}{S\arabic{equation}}
\renewcommand{\thefigure}{S\arabic{figure}}
\renewcommand{\thetable}{S\arabic{table}}
\renewcommand{\bibnumfmt}[1]{[S#1]}
\renewcommand{\citenumfont}[1]{S#1}

\section{Mean Field Hamiltonian for the Excitonic Insulating State}
In monolayer WTe$_2$, its valence band dispersion $\epsilon_{\textrm{v}}\left(\bm{k}\right)$ and conduction band dispersions $\epsilon_\pm\left(\bm{k}\right)$ can be approximated up to the quadratic order of wave vector:
\begin{align}\label{band_dispersion}
\epsilon_{\textrm{v}}\left(\bm{k}\right)\approx-\frac{\hbar^2k_x^2}{2m_{\textrm{v}, x}}-\frac{\hbar^2k_y^2}{2m_{\textrm{v}, y}}+\mu_{\textrm{v}},\quad \epsilon_\pm\left(\bm{k}\right)\approx\frac{\hbar^2k_x^2}{2m_{\textrm{c}, x}}+\frac{\hbar^2\left(k_y\mp q_0\right)^2}{2m_{\textrm{c}, y}}-\mu_{\textrm{c}}.
\end{align}
At Fermi energy, the electronic energy spectrum of monolayer WTe$_2$ generate a hole pocket centered at $\bm{k}=\bm{0}$ and two flanking electronic pockets centered at $\pm\bm{q}_0$ with $\bm{q}_0=\left(0, q_0\right)$. In the presence of Coulomb interaction, electronic states in the conduction bands tend to bind hole states from the valence band to form excitonic bound states, spontaneously generating density orders that gap out the whole Fermi surfaces. Due to the density order formed, the two electronic Fermi pockets are shifted toward $\bm{k}=\bm{0}$. In our consideration, the density order that couples the middle hole pocket and the two electronic pockets is assumed to be the pseudo-spin density order: $V_1\sim\sum_{\bm{k}}\left\langle \psi^\dagger_{+, \bm{k}+\bm{q}, \uparrow}\psi_{\textrm{v}, \bm{k}, \uparrow}-\psi^\dagger_{+, \bm{k}+\bm{q}, \downarrow}\psi_{\textrm{v}, \bm{k}, \downarrow} \right\rangle=\sum_{\bm{k}}\left\langle \psi^\dagger_{-, \bm{k}-\bm{q}, \uparrow}\psi_{\textrm{v}, \bm{k}, \uparrow}-\psi^\dagger_{-, \bm{k}-\bm{q}, \downarrow}\psi_{\textrm{v}, \bm{k}, \downarrow} \right\rangle$. The density order that hybridizes the two electronic Fermi pockets is assumed to be charge density order $V_2\sim\sum_{\bm{k}}\left\langle \psi^\dagger_{-, \bm{k}-\bm{q}, \uparrow}\psi_{+, \bm{k}+\bm{q}, \uparrow}+\psi^\dagger_{-, \bm{k}-\bm{q}, \downarrow}\psi_{+, \bm{k}+\bm{q}, \downarrow}\right\rangle$, where $\psi^{\left(\dagger\right)}_{\pm, \bm{k}, \uparrow/\downarrow}$ is to annihilate (create) a state at $\bm{k}$ in the band $\epsilon_{\pm}\left(\bm{k}\right)$ and $\psi^{\left(\dagger\right)}_{\textrm{v}, \bm{k}, \uparrow/\downarrow}$ is to annihilate (create) a state at $\bm{k}$ in the band $\epsilon_{\textrm{v}}\left(\bm{k}\right)$. The subscript $\uparrow/\downarrow$ is the pseudo-spin index. Recently, the pseudo-spin density order $V_1$ and charge density order $V_2$ have been confirmed in Hartree-Fock mean field calculation~\cite{Parameswaran_supp}. The mean field Hamiltonian for the excitonic insulating state has the form
\begin{align}\nonumber
\mathcal{H}=&\sum_{\bm{k}}\begin{pmatrix}
\psi^\dagger_{\textrm{v}, \bm{k}, \uparrow} & \psi^\dagger_{+, \bm{k}+\bm{q}, \uparrow} & \psi^\dagger_{-, \bm{k}-\bm{q}, \uparrow}
\end{pmatrix}\begin{pmatrix}
\epsilon_{\textrm{v}}\left(\bm{k}\right) & V_1 & V_1 \\
V_1 & \epsilon_+\left(\bm{k}+\bm{q}\right) & V_2 \\
V_1 & V_2 & \epsilon_-\left(\bm{k}-\bm{q}\right)
\end{pmatrix}\begin{pmatrix}
\psi_{\textrm{v}, \bm{k}, \uparrow} \\ \psi_{+, \bm{k}+\bm{q}, \uparrow} \\ \psi_{-, \bm{k}-\bm{q}, \uparrow}
\end{pmatrix}\\
&+\sum_{\bm{k}}\begin{pmatrix}
\psi^\dagger_{\textrm{v}, \bm{k}, \downarrow} & \psi^\dagger_{+, \bm{k}+\bm{q}, \downarrow} & \psi^\dagger_{-, \bm{k}-\bm{q}, \downarrow}
\end{pmatrix}\begin{pmatrix}
\epsilon_{\textrm{v}}\left(\bm{k}\right) & -V_1 & -V_1 \\
-V_1 & \epsilon_{+}\left(\bm{k}+\bm{q}\right) & V_2 \\
-V_1 & V_2 & \epsilon_{-}\left(\bm{k}-\bm{q}\right)
\end{pmatrix}\begin{pmatrix}
\psi_{\textrm{v}, \bm{k}, \downarrow} \\ \psi_{+, \bm{k}+\bm{q}, \downarrow} \\ \psi_{-, \bm{k}-\bm{q}, \downarrow}
\end{pmatrix}.
\end{align}
As the Hamiltonian matrix in the pseudo-spin up and pseudo-spin down sectors give the same energy eigenvalue, we can drop the pseudo-spin index for simplicity. Then the simplified mean field Hamiltonian takes the form
\begin{align}\label{H_M}
\mathcal{H}=\begin{pmatrix}
\psi^\dagger_{\textrm{v}, \bm{k}} & \psi^\dagger_{+, \bm{k}+\bm{q}} & \psi^\dagger_{-, \bm{k}-\bm{q}}
\end{pmatrix}\begin{pmatrix}
\epsilon_{\textrm{v}}\left(\bm{k}\right) & V_1 & V_1 \\
V_1 & \epsilon_{+}\left(\bm{k}+\bm{q}\right) & V_2 \\
V_1 & V_2 & \epsilon_{-}\left(\bm{k}-\bm{q}\right)
\end{pmatrix}\begin{pmatrix}
\psi_{\textrm{v}, \bm{k}} \\ \psi_{+, \bm{k}+\bm{q}} \\ \psi_{-, \bm{k}-\bm{q}}
\end{pmatrix},
\end{align}
where the density order $V_1$ can be either positive or negative. In order to figure out how the three bands couple through the density order $V_1$, $V_2$, we further introduce the unitary transofrmation to the Hamiltonian matrix
\begin{align}\nonumber
\tilde{H}=&\begin{pmatrix}
1 & 0 & 0 \\
0 & \cos\frac{\theta}{2} & \sin\frac{\theta}{2} \\
0 & \sin\frac{\theta}{2} & -\cos\frac{\theta}{2}
\end{pmatrix}\begin{pmatrix}
\epsilon_{\textrm{v}}\left(\bm{k}\right) & V_1 & V_1 \\
V_1 & \epsilon_{+}\left(\bm{k}+\bm{q}\right) & V_2 \\
V_1 & V_2 & \epsilon_{-}\left(\bm{k}-\bm{q}\right)
\end{pmatrix}\begin{pmatrix}
1 & 0 & 0 \\
0 & \cos\frac{\theta}{2} & \sin\frac{\theta}{2} \\
0 & \sin\frac{\theta}{2} & -\cos\frac{\theta}{2}
\end{pmatrix}\\
=&\left(\begin{smallmatrix}
-\frac{\hbar^2k_x^2}{2m_{\textrm{v}, x}}-\frac{\hbar^2k_y^2}{2m_{\textrm{v}, y}}+\mu_{\textrm{v}} & V_1\left(\cos\frac{\theta}{2}+\sin\frac{\theta}{2}\right) & V_1\left(-\cos\frac{\theta}{2}+\sin\frac{\theta}{2}\right) \\
V_1\left(\cos\frac{\theta}{2}+\sin\frac{\theta}{2}\right) & \frac{\hbar^2k_x^2}{2m_{\textrm{c}, x}}+\frac{\hbar^2\left[k_y^2+\left(q-q_0\right)^2\right]}{2m_{\textrm{c}, y}}-\mu_{\textrm{c}}+\sqrt{\left[\frac{\hbar^2k_y\left(q-q_0\right)}{m_{\textrm{c}, y}}\right]^2+V_2^2} & 0 \\
V_1\left(-\cos\frac{\theta}{2}+\sin\frac{\theta}{2}\right) & 0 & \frac{\hbar^2k_x^2}{2m_{\textrm{c}, x}}+\frac{\hbar^2\left[k_y^2+\left(q-q_0\right)^2\right]}{2m_{\textrm{c}, y}}-\mu_{\textrm{c}}-\sqrt{\left[\frac{\hbar^2k_y\left(q-q_0\right)}{m_{\textrm{c}, y}}\right]^2+V_2^2}
\end{smallmatrix}\right),
\end{align}
with
\begin{align}
\cos\theta=\frac{\hbar^2}{m_{\textrm{c}, y}}\frac{k_y\left(q-q_0\right)}{\sqrt{\left[\frac{\hbar^2k_y\left(q-q_0\right)}{m_{\textrm{c}, y}}\right]^2+V_2^2}},\quad\sin\theta=\frac{V_2}{\sqrt{\left[\frac{\hbar^2k_y\left(q-q_0\right)}{m_{\textrm{c}, y}}\right]^2+V_2^2}}.
\end{align}
After the unitary transformation, the conduction bands coupling sector is diagonalized in the Hamiltonian. The resulting new conduction bands are separated in energy, and couple individually with the valence band $\epsilon_{\textrm{v}}\left(\bm{k}\right)$. Importantly, one can find that $|-\cos\frac{\theta}{2}+\sin\frac{\theta}{2}|>|\cos\frac{\theta}{2}+\sin\frac{\theta}{2}|$ when $V_2<0$, which means that the coupling between the valence band $\epsilon_{\textrm{v}}\left(\bm{k}\right)$ and the lower energy conduction band $\frac{\hbar^2k_x^2}{2m_{\textrm{c}, x}}+\frac{\hbar^2\left[k_y^2+\left(q-q_0\right)^2\right]}{2m_{\textrm{c}, y}}-\mu_{\textrm{c}}-\sqrt{\left[\frac{\hbar^2k_y\left(q-q_0\right)}{m_{\textrm{c}, y}}\right]^2+V_2^2}$ is dominant. Since the insulating gap mainly arises from the hybridization between $\epsilon_{\textrm{v}}\left(\bm{k}\right)$ and $\frac{\hbar^2k_x^2}{2m_{\textrm{c}, x}}+\frac{\hbar^2\left[k_y^2+\left(q-q_0\right)^2\right]}{2m_{\textrm{c}, y}}-\mu_{\textrm{c}}-\sqrt{\left[\frac{\hbar^2k_y\left(q-q_0\right)}{m_{\textrm{c}, y}}\right]^2+V_2^2}$, $V_2<0$ is supposed to be energetically more favorable.

\section{Ground State Energy Minimization}
\begin{figure}
\centering
\includegraphics[width=3in]{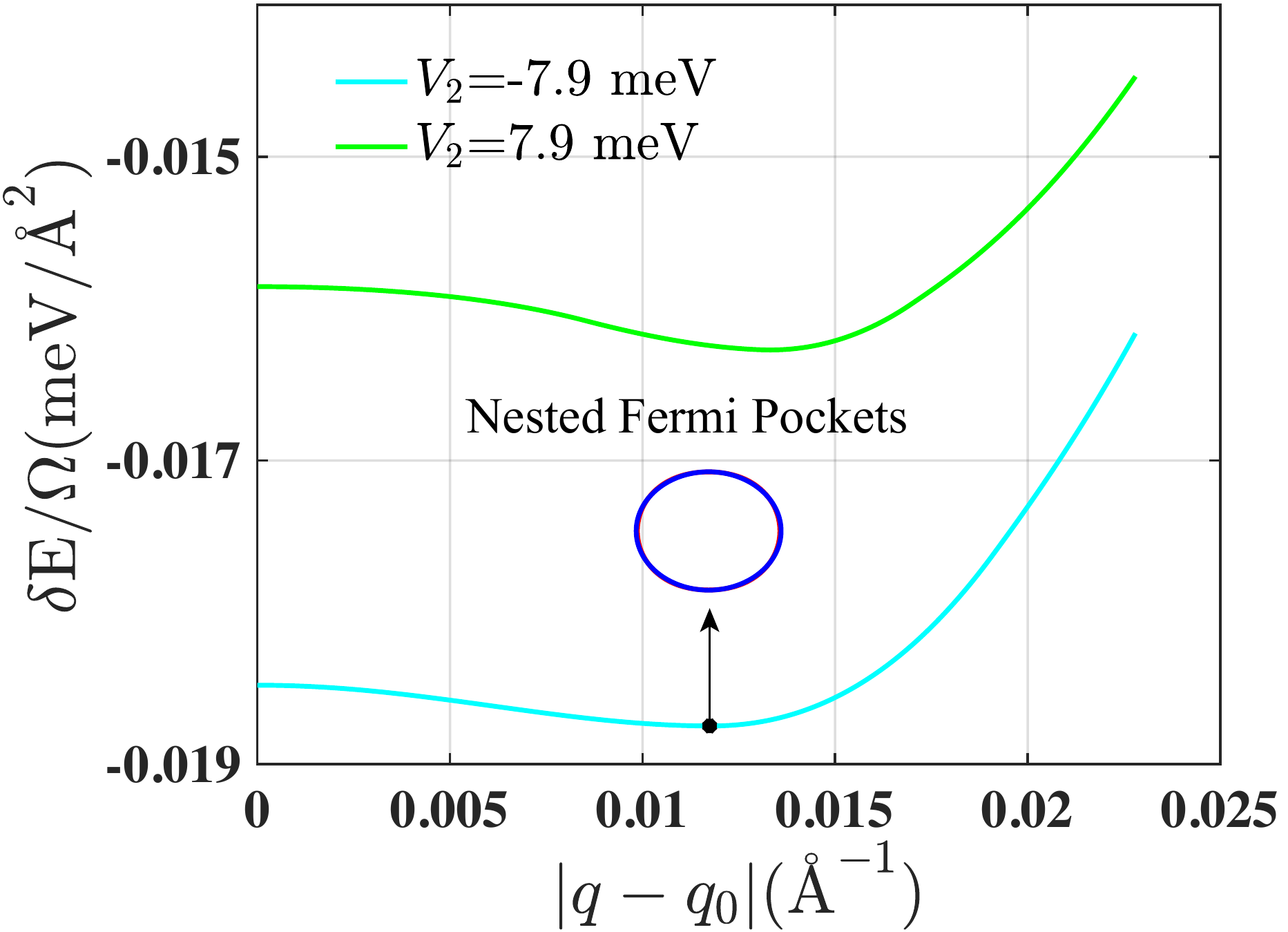}
\caption{The gained energy density as a function of $|\bm{q}-\bm{q}_0|$ at given coupling potential $\left(V_1,  V_2\right)=\left(1.3, \pm 7.9\right)$ meV. The ground state takes lower energy with negative coupling $V_2$. At the optimal shift momentum $\bm{q}$, the Fermi pockets with $\left(V_1, V_2\right)=\left(0, -7.9\right)$ meV is plotted. The blue colored pocket is from hybridized conduction bands, and the red pocket is from the valence band. The hybridized electronic Fermi pockets is well nested with the hole Fermi pocket. Band dispersions with $m_{\textrm{c}, x}=m_{\textrm{c}, y}=0.29m_{\textrm{e}}$, $m_{\textrm{v}, x}=\frac{2}{3}m_{\textrm{v}, y}=0.56m_{\textrm{e}}$ are taken in the calculations. The chemical potentials are $\mu_{\textrm{c}}=9.7$ meV, $\mu_{\textrm{v}}=8.1$ meV, which make the covered Fermi surface area $S$ give the oscillation frequency $f=\frac{\hbar S}{2\pi e}=48.6$ T.}\label{figS1}
\end{figure}

The ground state energy for the excitonic insulating state is
\begin{align}
\left\langle E \right\rangle=\frac{1}{\Omega}\sum_{\nu, \bm{k}}f\left[E_{\nu}\left(\bm{k}\right)\right]E_{\nu}\left(\bm{k}\right),
\end{align}
with $\Omega$ being the sample area, $f\left(\epsilon\right)=\frac{1}{2}\left(1-\tanh\frac{1}{2}\beta\epsilon\right)$ being the Fermi Dirac distribution, and $E_{\nu}\left(\bm{k}\right)$ being the energy eigenvalue of the mean field Hamiltonian matrix in Eq. \ref{H_M}. The ground state energy at zero Coulomb interaction can be obtained at $V_1=V_2=0$:
\begin{align}
\left\langle E \right\rangle_0=&\frac{1}{\Omega}\sum_{\nu, \bm{k}}\left\{f\left[\epsilon_{\textrm{v}}\left(\bm{k}\right)\right]\epsilon_{\textrm{v}}\left(\bm{k}+\bm{q}\right)+f\left[\epsilon_{+}\left(\bm{k}+\bm{q}\right)\right]\epsilon_{+}\left(\bm{k}\right)+f\left[\epsilon_{-}\left(\bm{k}-\bm{q}\right)\right]\epsilon_{-}\left(\bm{k}-\bm{q}\right)\right\}.
\end{align}
We know that opening an excitonic insulating gap reduces the ground state energy by $\delta E=\left\langle E \right\rangle-\left\langle E_0 \right\rangle$. By varying the shift momentum $\bm{q}$, the optimal $\bm{q}$ that minimizes the gained energy $\delta E$ can be found. In Fig. \ref{figS1}, the gained energy is calculated at $\left(V_1, V_2\right)=\left(1.3, \pm 7.9\right)$ meV. It is clear that the one with negative $V_2$ saves more energy. At the optimal $\bm{q}$, the hybridized electronic Fermi pockets and hole Fermi pocket, which is obtained from the energy contour at $V_1=0$ but $V_2=-7.9$ meV, is plotted as the inset of Fig. \ref{figS1}. The hybridized electronic Fermi pockets are well nested with the hole Fermi pocket.

\section{Landau Levels Spectrum Calculation}
\subsection{Landau Levels from the Valence and Conduction Bands}
The valence band dispersion $\epsilon_{\textrm{v}}\left(\bm{k}\right)$ and the conduction band dispersions $\epsilon_{\pm}\left(\bm{k}\pm\bm{q}\right)$ in Eq. \ref{band_dispersion} can go back to the form of Schr\"odinger equation after the substituting $\bm{k}\rightarrow-i\bm{\nabla}$:
\begin{align}\label{Sc_v}
\left(\frac{\partial^2}{2m_{\textrm{v}, x}}\frac{\partial^2}{\partial x^2}+\frac{\hbar^2}{2m_{\textrm{v}, y}}\frac{\partial^2}{\partial y^2}+\mu_{\textrm{v}}\right)\psi_{\textrm{v}}\left(x, y\right)=&\epsilon_{\textrm{v}}\psi_{\textrm{v}}\left(x, y\right),\\\label{Sc_c}
\left\{-\frac{\hbar^2}{2m_{\textrm{c}, x}}\frac{\partial^2}{\partial x^2}-\frac{\hbar^2}{2m_{\textrm{c}, y}}\left[\frac{\partial}{\partial y}\pm i\left(q-q_0\right)\right]^2-\mu_{\textrm{c}}\right\}\psi_\pm\left(x, y\right)=&\epsilon_\pm\psi_\pm\left(x, y\right).
\end{align}
In the presence of magnetic field $\bm{B}=\left(0, 0, B\right)$, we use the Landau gauge $\bm{A}=\left(0, Bx, 0\right)$ and in Eq. \ref{Sc_v}, Eq. \ref{Sc_c} do the substitution $-i\hbar\bm{\nabla}\rightarrow-i\hbar\bm{\nabla}-e\bm{A}$. Then the Schr\"odinger type equations become
\begin{align}\label{Sc_v2}
\left[\frac{\hbar^2}{2m_{\textrm{v}, x}}\frac{\partial^2}{\partial x^2}+\frac{\hbar^2}{2m_{\textrm{v}, y}}\left(\frac{\partial}{\partial y}-\frac{ieBx}{\hbar}\right)^2+\mu_{\textrm{v}}\right]\psi_{\textrm{v}}\left(x\right)=&\epsilon_{\textrm{v}}\psi_{\textrm{v}}\left(x\right),\\\label{Sc_c2}
\left\{-\frac{\hbar^2}{2m_{\textrm{c}, x}}\frac{\partial^2}{\partial x^2}-\frac{\hbar^2}{2m_{\textrm{c}, y}}\left[\frac{\partial}{\partial y}\pm i\left(q-q_0\right)-\frac{ieBx}{\hbar}\right]^2-\mu_{\textrm{c}}\right\}\psi_{\pm}\left(x\right)=&\epsilon_\pm\psi_{\pm}\left(x\right).
\end{align}
Then by substituting $\psi_{\textrm{v}}\left(x, y\right)=\phi_{\textrm{v}}\left(x\right)e^{ik_yy}$ and $\psi_\pm\left(x, y\right)=\phi_\pm\left(x\right)e^{ik_yy}$ into Eq. \ref{Sc_v2}, Eq. \ref{Sc_c2}, we can obtain
\begin{align}\label{Sc_v3}
-\left[-\frac{\hbar^2}{2m_{\textrm{v}, x}}\frac{\partial^2}{\partial x^2}+\frac{1}{2}m_{\textrm{v}, x}\omega^2_{\textrm{v}}\left(x-x_0\right)^2-\mu_{\textrm{v}}\right]\phi_{\textrm{c}}\left(x\right)=&\epsilon_{\textrm{v}}\phi_{\textrm{v}}\left(x\right),\\\label{Sc_c3}
\left\{-\frac{\hbar^2}{2m_{\textrm{c}, x}}\frac{\partial^2}{\partial x^2}+\frac{1}{2}m_{\textrm{c}, x}\omega^2_{\textrm{c}}\left[x\mp\frac{\hbar\left(q-q_0\right)}{eB}-x_0\right]^2-\mu_{\textrm{c}}\right\}\phi_\pm\left(x\right)=&\epsilon_\pm\phi_\pm\left(x\right),
\end{align}
with $\omega_{\textrm{v}}=\frac{eB}{m_{\textrm{v}}}$, $\omega_{\textrm{c}}=\frac{eB}{m_{\textrm{v}}}$ and $x_0=\frac{\hbar}{eB}$. Here the cyclotron masses are $m_{\textrm{v}}=\sqrt{m_{\textrm{v}, x}m_{\textrm{v}, y}}$, $m_{\textrm{v}}=\sqrt{m_{\textrm{c}, x}m_{\textrm{c}, y}}$. It is clear that the equations in Eq. \ref{Sc_v3}, Eq. \ref{Sc_c3} describe the simple harmonic oscillator, so the energy eigenvalues are
\begin{align}\label{Eq_LL}
\epsilon_{\textrm{v}, n}=-\left(n+\frac{1}{2}\right)\hbar\omega_{\textrm{v}}+\mu_{\textrm{v}},\quad \epsilon_{\pm, n}=\left(n+\frac{1}{2}\right)\hbar\omega_{\textrm{c}}-\mu_{\textrm{c}},\quad\textrm{with}\quad n\in\mathbb{Z}.
\end{align}

\subsection{Hamiltonian in Terms of Creation and Annihilation Operators}
For the simple harmonic oscillator type eigen equations in Eq. \ref{Sc_v3}, Eq. \ref{Sc_c3}, we can define the annihilation and creation operators to be
\begin{align}
a_{\textrm{v}}=&\sqrt{\frac{m_{\textrm{v}, x}\omega_{\textrm{v}}}{2\hbar}}\left(x-x_0\right)+\sqrt{\frac{\hbar}{2m_{\textrm{v}, x}\omega_{\textrm{v}}}}\frac{\partial}{\partial x},\quad a^\dagger_{\textrm{v}}=\sqrt{\frac{m_{\textrm{v}, x}\omega_{\textrm{v}}}{2\hbar}}\left(x-x_0\right)-\sqrt{\frac{\hbar}{2m_{\textrm{v}, x}\omega_{\textrm{v}}}}\frac{\partial}{\partial x},\\
a_{\textrm{c}}=&\sqrt{\frac{m_{\textrm{c}, x}\omega_{\textrm{c}}}{2\hbar}}\left(x-x_0\right)+\sqrt{\frac{\hbar}{2m_{\textrm{c}, x}\omega_{\textrm{c}}}}\frac{\partial}{\partial x},\quad a^\dagger_{\textrm{c}}=\sqrt{\frac{m_{\textrm{c}, x}\omega_{\textrm{c}}}{2\hbar}}\left(x-x_0\right)-\sqrt{\frac{\hbar}{2m_{\textrm{c}, x}\omega_{\textrm{c}}}}\frac{\partial}{\partial x}.
\end{align}
Then the Hamiltonians in Eq. \ref{Sc_v3} and Eq. \ref{Sc_c3} can be expressed in terms of the annihilation and creation operators as
\begin{align}
\hat{H}_{\textrm{v}}=&\frac{\hbar^2}{2m_{\textrm{v}, x}}\frac{\partial^2}{\partial x^2}-\frac{1}{2}m_{\textrm{v}, x}\omega^2_{\textrm{v}}\left(x-x_0\right)^2+\mu_{\textrm{v}}=-\frac{1}{2}\hbar\omega_{\textrm{v}}\left(a^\dagger_{\textrm{v}}a_{\textrm{v}}+a_{\textrm{v}}a^\dagger_{\textrm{v}}\right)+\mu_{\textrm{v}},\\\nonumber
\hat{H}_\pm=&-\frac{\hbar^2}{2m_{\textrm{c}, x}}\frac{\partial^2}{\partial x^2}+\frac{1}{2}m_{\textrm{c}, x}\omega^2_{\textrm{c}}\left[x\mp\frac{\hbar\left(q-q_0\right)}{eB}-x_0\right]^2-\mu_{\textrm{c}}\\
=&\frac{1}{2}\hbar\omega_{\textrm{c}}\left[a^\dagger_{\textrm{c}}\mp\frac{l_{\textrm{c}, B}}{\sqrt{2}}\left(q-q_0\right)\right]\left[a_{\textrm{c}}\mp\frac{l_{\textrm{c}, B}}{\sqrt{2}}\left(q-q_0\right)\right]+\frac{1}{2}\hbar\omega_{\textrm{c}}\left[a_{\textrm{c}}\mp\frac{l_{\textrm{c}, B}}{\sqrt{2}}\left(q-q_0\right)\right]\left[a^\dagger_{\textrm{c}}\mp\frac{l_{\textrm{c}, B}}{\sqrt{2}}\left(q-q_0\right)\right]-\mu_{\textrm{c}},
\end{align}
with the magnetic length for the electronic states being $l_{\textrm{c}, B}=\sqrt{\frac{\hbar}{m_{\textrm{c}, x}\omega_{\textrm{c}}}}=\left(\frac{m_{\textrm{c}, y}}{m_{\textrm{c}, x}}\right)^{\frac{1}{4}}\sqrt{\frac{\hbar}{eB}}$. The hole occupation state and electronic occupation state that are associated with the creation operators $a^\dagger_{\textrm{v}}$, $a^\dagger_{\textrm{c}}$ can be further defined as
\begin{align}
\ket{\textrm{v}, n}=&\frac{\left(a^\dagger_{\textrm{v}}\right)^n}{\sqrt{n!}}\ket{\textrm{v},  0}=\frac{1}{\sqrt{n!}}\left[\sqrt{\frac{m_{\textrm{v}, x}\omega_{\textrm{v}}}{2\hbar}}\left(x-x_0\right)-\sqrt{\frac{\hbar}{2m_{\textrm{v}, x}\omega_{\textrm{v}}}}\frac{\partial}{\partial x}\right]^n\phi_{\textrm{v}, 0}\left(x\right)=\phi_{\textrm{v}, n}\left(x\right),\\
\ket{\textrm{c}, n}=&\frac{\left(a^\dagger_{\textrm{c}}\right)^n}{\sqrt{n!}}\ket{\textrm{c},  0}=\frac{1}{\sqrt{n!}}\left[\sqrt{\frac{m_{\textrm{c}, x}\omega_{\textrm{c}}}{2\hbar}}\left(x-x_0\right)-\sqrt{\frac{\hbar}{2m_{\textrm{c}, x}\omega_{\textrm{c}}}}\frac{\partial}{\partial x}\right]^n\phi_{\textrm{c}, 0}\left(x\right)=\phi_{\textrm{c}, n}\left(x\right),
\end{align}
where $\phi_{\textrm{v}, n}\left(x\right)$ and $\phi_{\textrm{c}, n}\left(x\right)$ are
\begin{align}
\phi_{\textrm{v}, n}\left(x\right)=&\frac{1}{\pi^{\frac{1}{4}}l^{\frac{1}{2}}_{\textrm{v}, B}\sqrt{2^nn!}}H_n\left(\frac{x-x_0}{l_{\textrm{v}, B}}\right)e^{-\frac{\left(x-x_0\right)^2}{2l^2_{\textrm{v}, B}}},\quad\textrm{with}\quad l_{\textrm{v}, B}=\sqrt{\frac{\hbar}{m_{\textrm{v}, x}\omega_{\textrm{v}}}}=\left(\frac{m_{\textrm{v}, y}}{m_{\textrm{v}, x}}\right)^{\frac{1}{4}}\sqrt{\frac{\hbar}{eB}},\\
\phi_{\textrm{c}, n}\left(x\right)=&\frac{1}{\pi^{\frac{1}{4}}l^{\frac{1}{2}}_{\textrm{c}, B}\sqrt{2^nn!}}H_n\left(\frac{x-x_0}{l_{\textrm{c}, B}}\right)e^{-\frac{\left(x-x_0\right)^2}{2l^2_{\textrm{c}, B}}}.
\end{align}
Here $H_n\left(x\right)$ is the usual Hermite function. Importantly, the matrix representation for the creation and annihilation operators can be obtained in the occupation basis as
\begin{align}
\bra{\textrm{v}, n} a_{\textrm{v}} \ket{\textrm{v}, m}=\bra{\textrm{c}, n} a_{\textrm{c}} \ket{\textrm{c}, m}=\begin{pmatrix}
0 & 1 & 0 & 0 & \dots & 0 \\
0 & 0 & \sqrt{2} & 0 & \dots & 0 \\
0 & 0 & 0 & \sqrt{3} & \dots & 0 \\
0 & 0 & 0 & 0 & \ddots & 0 \\
\vdots & \vdots  & \vdots & \vdots & \ddots & \sqrt{n} \\
0 & 0 & 0 & 0 & \dots & 0
\end{pmatrix},\quad\bra{\textrm{v}, n}a^\dagger_{\textrm{v}}\ket{\textrm{v}, m}=\bra{\textrm{c}, n}a_{\textrm{c}}^\dagger\ket{\textrm{c}, m}=\begin{pmatrix}
0 & 0 & 0 & 0 & \dots & 0 \\
1 & 0 & 0 & 0 & \dots & 0 \\
0 & \sqrt{2} & 0 & 0 & \dots & 0 \\
0 & 0 & \sqrt{3} & 0 & \dots & 0 \\
\vdots & \vdots  & \vdots & \ddots & \ddots & 0 \\
0 & 0 & 0 & 0 & \sqrt{n} & 0
\end{pmatrix}.
\end{align}
In the occupation basis, the Hamiltonian $\hat{H}_{\textrm{v}}$ and $\hat{H}_\pm$ can have the matrix form. Diagonalizing the Hamiltonian matrix of $\hat{H}_{\textrm{v}}$, $\hat{H}_\pm$ yields the energy eigenvalues $\epsilon_{\textrm{v}, n}=-\left(n+\frac{1}{2}\right)\hbar\omega_{\textrm{v}}+\mu_{\textrm{v}}$, $\epsilon_{\pm, n}=\left(n+\frac{1}{2}\right)\hbar\omega_{\textrm{c}}-\mu_{\textrm{c}}$ with $n=0, 1, 2, 3, \dots$.

\subsection{Landau Levels Hybridization}
The Hamiltonian that describes the hybridization of Landau levels takes similar form as the mean field Hamiltonian in Eq. \ref{H_M}
\begin{align}\label{H_hat}
\hat{H}=\begin{pmatrix}
\hat{H}_{\textrm{v}} & \hat{V}_1 & \hat{V}_1 \\
\hat{V}^\dagger_1 & \hat{H}_+ & \hat{V}_2 \\
\hat{V}_1^\dagger & \hat{V}^\dagger_2 & \hat{H}_-
\end{pmatrix}.
\end{align}
In the occupation basis, all the matrix elements of the Hamiltonian $\hat{H}$ can be calculated. The matrix elements in $\hat{H}_{\textrm{v}}$, $\hat{H}_+$ and $\hat{H}_-$ can be obtained through the matrix representation of the creation and annihilation operator $a^\dagger_{\textrm{v}}$, $a_{\textrm{v}}$, $a^\dagger_{\textrm{c}}$ and $a_{\textrm{c}}$. The matrix elements of $\hat{V}_1$, $\hat{V}_2$ are obtained through integral
\begin{align}
\hat{V}_{1, n, m}=&\bra{\textrm{v}, n}V_1\ket{\textrm{c}, m}=\frac{V_1}{\sqrt{\pi l_{\textrm{v}, B}l_{\textrm{c}, B}2^{n+m}n!m!}}\int_{-\infty}^\infty H_n\left(\frac{x-x_0}{l_{\textrm{v}, B}}\right)H_m\left(\frac{x-x_0}{l_{\textrm{c}, B}}\right)e^{-\frac{1}{2}\left(\frac{1}{l^2_{\textrm{v}, B}}+\frac{1}{l^2_{\textrm{c}, B}}\right)\left(x-x_0\right)^2}dx,\\
\hat{V}_{2, n, m}=&\bra{\textrm{c}, n}V_2\ket{\textrm{c}, m}=V_2\delta_{n, m},
\end{align}
where $\delta_{n, m}$ is Kronecker Delta function. In order to calculate the matrix elements in $\hat{V}_1$, we need to introduce the integral formula~\cite{Quattromini}:
\begin{align}
\int_{-\infty}^\infty H_n\left(ax+b\right)H_m\left(cx+d\right)e^{-fx^2+\alpha x}dx=\sqrt{\frac{\pi}{f}}e^{\frac{\alpha^2}{4f}}\tilde{H}_{n, m}\left(ab+\frac{a}{f}\alpha, -1+\frac{a^2}{f}; 2d+\frac{c}{f}\alpha, -1+\frac{c^2}{f} | \frac{2ac}{f}\right).
\end{align}
Here $\tilde{H}_{n, m}\left(x, y; w, z | \tau\right)$ is the two-index Hermite function
\begin{align}
\tilde{H}_{n, m}\left(x, y; w, z | \tau\right)=\sum_{k=0}^{\textrm{min}\left(m, n\right)}\frac{n!m!}{\left(n-k\right)!\left(m-k\right)!k!}\tau^k\tilde{H}_{n-k}\left(x, y\right)\tilde{H}_{m-k}\left(w, z\right),
\end{align}
where $\tilde{H}_n\left(x ,y\right)$ is the two variable Hermite function $\tilde{H}_n\left(x, y\right)=n!\sum_{k=0}^{\left[\frac{n}{2}\right]}\frac{x^{n-2k}y^k}{\left(n-2k\right)!k!}$. As a result, we can then get the matrix elements in $\hat{V}_1$ to be
\begin{align}\nonumber\label{Coupling_V1}
\hat{V}_{1, n, m}=&\frac{V_1}{\sqrt{2^{n+m-1}n!m!\left(\frac{l_{\textrm{v}, B}}{l_{\textrm{c}, B}}+\frac{l_{\textrm{c}, B}}{l_{\textrm{v}, B}}\right)}}\tilde{H}_{n, m}\left(0, -\frac{l^2_{\textrm{v}, B}-l^2_{\textrm{c}, B}}{l^2_{\textrm{v}, B}+l^2_{\textrm{c}, B}}; 0, \frac{l^2_{\textrm{v}, B}-l^2_{\textrm{c}, B}}{l^2_{\textrm{v}, B}+l^2_{\textrm{c}, B}}|\frac{4l_{\textrm{v}, B}l_{\textrm{c}, B}}{l^2_{\textrm{v}, B}+l^2_{\textrm{c}, B}}\right)\\
=&\frac{V_1}{\sqrt{2^{n+m-1}n!m!\left(\gamma^{-\frac{1}{4}}+\gamma^{\frac{1}{4}}\right)}}\tilde{H}_{n, m}\left(0, \frac{1-\sqrt{\gamma}}{1+\sqrt{\gamma}}; 0, -\frac{1-\sqrt{\gamma}}{1+\sqrt{\gamma}} | \frac{4}{\gamma^{-\frac{1}{4}}+\gamma^{\frac{1}{4}}}\right).
\end{align}
Here we have defined the relative anisotropy $\gamma=\frac{m_{\textrm{v}, y}m_{\textrm{c}, x}}{m_{\textrm{v}, x}m_{\textrm{c}, y}}$. Now in the occupation basis all the matrix elements of $\hat{H}$ in Eq. \ref{H_hat} can be calculated, so the energy eigenvalue of $\hat{H}$ is obtained through numerical diagonalization of the matrix.

\begin{figure}
\centering
\includegraphics[width=6.8in]{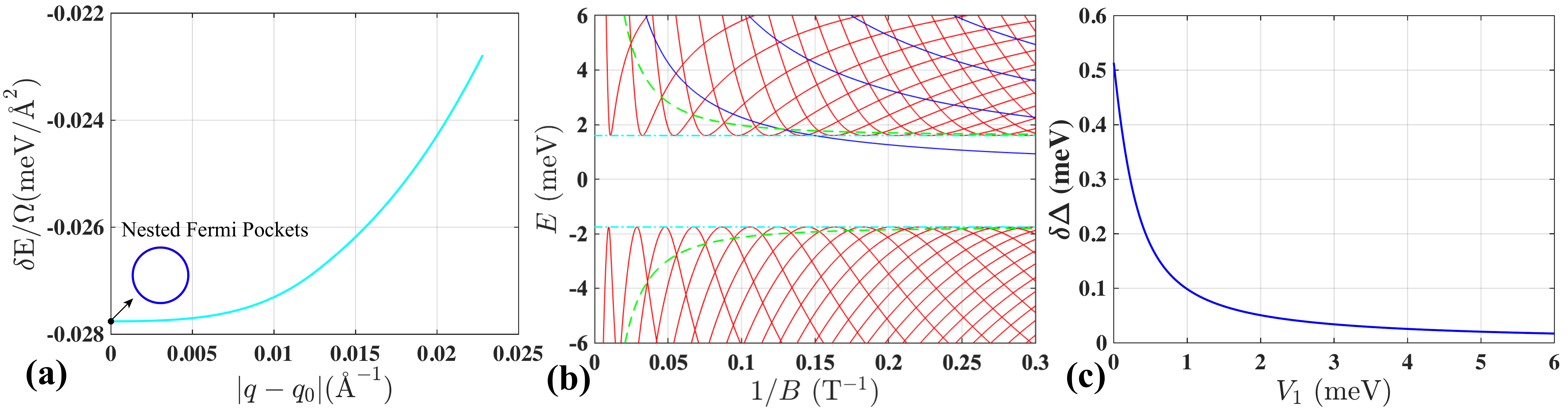}
\caption{(a) The gained energy density for the case of $\gamma=1$ at given coupling potential $\left(V_1, V_2\right)=\left(1.3, -10\right)$ meV. Since there is no relative anisotropy between the electronic and hole Fermi pockets, the ground state energy is minimized at the optimal shift momentum $\bm{q}=\bm{q}_0$. The inset is the nested Fermi pockets at $\left(V_1, V_2\right)=\left(0, -10\right)$ meV. (b) The Landau levels spectrum for $\gamma=1$ case. The coupling potential is the same as that in (a). The red lines are the spectrum from $E_{1, n}$, $E_{2, n}$. The blue lines are from $E_{3, n}$. The green dashed lines correspond to $E_{1, \textrm{cro}}$, $E_{2, \textrm{cro}}$, and the cyan lines denote the boundary of the gap at $E_{1, \textrm{min}}$, $E_{2, \textrm{max}}$. (c) The gap modulation $\delta\Delta$ at $1/B=0.23$ T$^{-1}$. In all the calculations, cyclotron masses are taken to be $m_{\textrm{c}}=0.29m_{\textrm{e}}$, $m_{\textrm{v}}=0.67m_{\textrm{e}}$. The chemical potentials set to be $\mu_{\textrm{c}}=9.7$ meV, $\mu_{\textrm{v}}=8.1$ so that the covered Fermi surface area $S$ gives the oscillation frequency $f=\frac{\hbar S}{2\pi e}=48.6$ T.}\label{figS2}
\end{figure}

\section{Analytical Results for the Landau Levels Spectrum in the Case of $\gamma=1$}
When the relative anisotropy takes $\gamma=1$, the electronic Fermi pockets and the hole Fermi pocket are in the same shape, so at $\bm{q}=\bm{q}_0$ the hybridized electronic Fermi pocket can be perfectly nested with the hole Fermi pocket as seen in the inset of Fig. \ref{figS2} (a). The perfect nesting at $\bm{q}=\bm{q}_0$ is confirmed to be the minimal ground state energy in the case of $\gamma=1$, as is shown in Fig. \ref{figS2}. There the two electronic pockets hybridizes at $V_2=-10$ meV and only one Fermi circle centered at $\bm{k}=\bm{0}$ is left to further couple to the hole Fermi pocket. At $\gamma=1$, the coupling matrix $\hat{V}_1$ is found to be $\hat{V}_1=V_1I$ with $I$ being the identity matrix. The diagonal form of both $\hat{V}_1$, $\hat{V}_2$ means that the Landau levels only couple with levels of the same index, so the Hamiltonian that describes the Landau levels takes the simple form
\begin{align}
H_{\textrm{LL}}=&\begin{pmatrix}
-\left(n+\frac{1}{2}\right)\hbar\omega_{\textrm{v}}+\mu_{\textrm{v}} & V_1 & V_1 \\
V_1 & \left(n+\frac{1}{2}\right)\hbar\omega_{\textrm{c}}-\mu_{\textrm{c}} & V_2 \\
V_1 & V_2 & \left(n+\frac{1}{2}\right)\hbar\omega_{\textrm{c}}-\mu_{\textrm{c}}
\end{pmatrix}.
\end{align}
After a unitary transformation, the Hamiltonian matrix $H_{\textrm{LL}}$ can be block diagonalized to be
\begin{align}
\tilde{H}_{\textrm{LL}}=&\begin{pmatrix}
-\left(n+\frac{1}{2}\right)\hbar\omega_{\textrm{v}}+\mu_{\textrm{v}} & \sqrt{2}V_1 & 0 \\
\sqrt{2}V_1 & \left(n+\frac{1}{2}\right)\hbar\omega_{\textrm{c}}-\mu_{\textrm{c}}+V_2 & 0 \\
0 & 0 & \left(n+\frac{1}{2}\right)\hbar\omega_{\textrm{c}}-\mu_{\textrm{c}}-V_2
\end{pmatrix}.
\end{align}
The energy eigenvalues of the matrix $\tilde{H}_{\textrm{LL}}$ can be analytically solved to be
\begin{align}
E_{1, n}=&\frac{1}{2}\left[\hbar\left(\omega_{\textrm{c}}-\omega_{\textrm{v}}\right)\left(n+\frac{1}{2}\right)-\mu_{\textrm{c}}+\mu_{\textrm{v}}+V_2\right]+\sqrt{\frac{1}{4}\left[\hbar\left(\omega_{\textrm{c}}+\omega_{\textrm{v}}\right)\left(n+\frac{1}{2}\right)-\mu_{\textrm{c}}-\mu_{\textrm{v}}+V_2\right]^2+2V_1^2},\\
E_{2, n}=&\frac{1}{2}\left[\hbar\left(\omega_{\textrm{c}}-\omega_{\textrm{v}}\right)\left(n+\frac{1}{2}\right)-\mu_{\textrm{c}}+\mu_{\textrm{v}}+V_2\right]-\sqrt{\frac{1}{4}\left[\hbar\left(\omega_{\textrm{c}}+\omega_{\textrm{v}}\right)\left(n+\frac{1}{2}\right)-\mu_{\textrm{c}}-\mu_{\textrm{v}}+V_2\right]^2+2V_1^2},\\
E_{3, n}=&\hbar\omega_{\textrm{c}}\left(n+\frac{1}{2}\right)-\mu_{\textrm{v}}-V_2.
\end{align}
The Landau levels spectrum from $E_{1, n}$,  $E_{2, n}$ and $E_{3, n}$ are plotted in Fig. \ref{figS2} (b). One can see that the Landau energy levels $-\left(n+\frac{1}{2}\right)\hbar\omega_{\textrm{v}}+\mu_{\textrm{v}}$ from the valence band couple with the Landau levels $\left(n+\frac{1}{2}\right)\hbar\omega_{\textrm{c}}-\mu_{\textrm{c}}+V_2$ from the conduction band. In this way, a hybridization gap arises and it is modulated by the applied magnetic field $B$. 

The gap modulation in the $\gamma=1$ case can also be analytically solved. At given $V_1$, $V_2$, the minimum of $E_{1, n}$ and the maximum of $E_{2, n}$ is
\begin{align}
E_{1, \textrm{min}}=\frac{\sqrt{m_{\textrm{c}}m_{\textrm{v}}}}{m_{\textrm{c}}+m_{\textrm{v}}}2\sqrt{2}V_1+\frac{m_{\textrm{v}}\mu_{\textrm{v}}-m_{\textrm{c}}\mu_{\textrm{c}}+m_{\textrm{c}}V_2}{m_{\textrm{c}}+m_{\textrm{v}}},\quad E_{2, \textrm{max}}=-\frac{\sqrt{m_{\textrm{c}}m_{\textrm{v}}}}{m_{\textrm{c}}+m_{\textrm{v}}}2\sqrt{2}V_1+\frac{m_{\textrm{v}}\mu_{\textrm{v}}-m_{\textrm{c}}\mu_{\textrm{c}}+m_{\textrm{c}}V_2}{m_{\textrm{c}}+m_{\textrm{v}}}.
\end{align}
The curve $E_{1, \textrm{cro}}$ that connects the lowest crossing points in $E_{1, n}$ and the curve $E_{2, \textrm{cro}}$ connecting the highest crossing points in $E_{2, n}$ are solved to be
\begin{align}\nonumber
E_{1, \textrm{cro}}=&\frac{1}{2}\left[\frac{m_{\textrm{c}}+m_{\textrm{v}}}{\sqrt{m_{\textrm{c}}m_{\textrm{v}}}}\sqrt{2V_1^2+\frac{\left(m_{\textrm{c}}^2+m_{\textrm{v}}^2\right)\hbar^2\omega_{\textrm{c}}\omega_{\textrm{v}}}{2\left(m_{\textrm{c}}+m_{\textrm{v}}\right)^2}+\frac{\sqrt{m_{\textrm{c}}m_{\textrm{v}}}\hbar\left(\omega_{\textrm{c}}-\omega_{\textrm{v}}\right)}{m_{\textrm{c}}+m_{\textrm{v}}}\sqrt{\frac{1}{4}\hbar^2\omega_{\textrm{c}}\omega_{\textrm{v}}+2V_1^2}}\right.\\
&\left.-\frac{\left(m_{\textrm{c}}-m_{\textrm{v}}\right)^2}{\sqrt{m_{\textrm{c}}m_{\textrm{v}}}\left(m_{\textrm{c}}+m_{\textrm{v}}\right)}\sqrt{\frac{1}{4}\hbar^2\omega_{\textrm{c}}\omega_{\textrm{v}}+2V_1^2}-2\hbar\left(\omega_{\textrm{c}}-\omega_{\textrm{v}}\right)\right]+\frac{m_{\textrm{v}}\mu_{\textrm{v}}-m_{\textrm{c}}\mu_{\textrm{c}}+m_{\textrm{c}}V_2}{m_{\textrm{c}}+m_{\textrm{v}}},\\\nonumber
E_{2, \textrm{cro}}=&\frac{1}{2}\left[-\frac{m_{\textrm{c}}+m_{\textrm{v}}}{\sqrt{m_{\textrm{c}}m_{\textrm{v}}}}\sqrt{2V_1^2+\frac{\left(m_{\textrm{c}}^2+m_{\textrm{v}}^2\right)\hbar^2\omega_{\textrm{c}}\omega_{\textrm{v}}}{2\left(m_{\textrm{c}}+m_{\textrm{v}}\right)^2}-\frac{\sqrt{m_{\textrm{c}}m_{\textrm{v}}}\hbar\left(\omega_{\textrm{c}}-\omega_{\textrm{v}}\right)}{m_{\textrm{c}}+m_{\textrm{v}}}\sqrt{\frac{1}{4}\hbar^2\omega_{\textrm{c}}\omega_{\textrm{v}}+2V_1^2}}\right.\\
&\left.+\frac{\left(m_{\textrm{c}}-m_{\textrm{v}}\right)^2}{\sqrt{m_{\textrm{c}}m_{\textrm{v}}}\left(m_{\textrm{c}}+m_{\textrm{v}}\right)}\sqrt{\frac{1}{4}\hbar^2\omega_{\textrm{c}}\omega_{\textrm{v}}+2V_1^2}-2\hbar\left(\omega_{\textrm{c}}-\omega_{\textrm{v}}\right)\right]+\frac{m_{\textrm{v}}\mu_{\textrm{v}}-m_{\textrm{c}}\mu_{\textrm{c}}+m_{\textrm{c}}V_2}{m_{\textrm{c}}+m_{\textrm{v}}}.
\end{align}
The curves $E_{1, \textrm{cro}}$ and $E_{2, \textrm{cro}}$ are plotted in Fig. \ref{figS2} (b). From $E_{1, \textrm{cro}}$, $E_{2, \textrm{cro}}$, $E_{1, \textrm{min}}$, and $E_{2, \textrm{max}}$, the amplitude of the energy oscillation at the gap boundary can then be obtained as
\begin{align}\nonumber\label{Gap_modulate_isotropy}
\delta\Delta=&E_{1, \textrm{cro}}-E_{1, \textrm{min}}=E_{2, \textrm{max}}-E_{2, \textrm{cro}}\\
=&\frac{1}{2}\left[\frac{m_{\textrm{c}}+m_{\textrm{v}}}{\sqrt{m_{\textrm{c}}m_{\textrm{v}}}}\sqrt{2V_1^2+\frac{\left(m_{\textrm{c}}^2+m_{\textrm{v}}^2\right)\hbar^2\omega_{\textrm{c}}\omega_{\textrm{v}}}{2\left(m_{\textrm{c}}+m_{\textrm{v}}\right)^2}+\frac{\sqrt{m_{\textrm{c}}m_{\textrm{v}}}\hbar\left(\omega_{\textrm{c}}-\omega_{\textrm{v}}\right)}{m_{\textrm{c}}+m_{\textrm{v}}}\sqrt{\frac{1}{4}\hbar^2\omega_{\textrm{c}}\omega_{\textrm{v}}+2V_1^2}}\right.\\
&\left.-\frac{\left(m_{\textrm{c}}-m_{\textrm{v}}\right)^2}{\sqrt{m_{\textrm{c}}m_{\textrm{v}}}\left(m_{\textrm{c}}+m_{\textrm{v}}\right)}\sqrt{\frac{1}{4}\hbar^2\omega_{\textrm{c}}\omega_{\textrm{v}}+2V_1^2}-2\hbar\left(\omega_{\textrm{c}}-\omega_{\textrm{v}}\right)\right]-\frac{\sqrt{m_{\textrm{c}}m_{\textrm{v}}}}{m_{\textrm{c}}+m_{\textrm{v}}}2\sqrt{2}V_1.
\end{align}
We know that the gap modulation is manifested by the amplitiude of the energy oscillation at the gap boundary, which is $\delta\Delta$. The amplitude of energy oscillation $\delta\Delta$ is determined by the cyclontron masses $m_{\textrm{v}}$, $m_{\textrm{c}}$, the hybridization $V_1$ and the applied magnetic field. In the limit $\hbar^2\omega_{\textrm{c}}\omega_{\textrm{c}}\ll V_1^2$, $\delta\Delta$ can be approximated as $\delta\Delta\approx\frac{\sqrt{m_{\textrm{c}}m_{\textrm{v}}}}{m_{\textrm{c}}+m_{\textrm{v}}}\frac{\hbar^2\omega_{\textrm{c}}\omega_{\textrm{v}}}{4\sqrt{2}V_1}$,  and one can see that the gap modulation gets larger in smaller cyclotron masses case. For a specific monolayer WTe$_2$-like material, the cyclotron masses are fixed, so $\delta\Delta$ is a function of $V_1$ at given magnetic field $1/B=0.23$ T$^{-1}$. In Fig. \ref{figS2} (c), $\delta\Delta$ is plotted in terms of $V_1$. It is clearly to see that $V_1<3.4$ meV delimits the region that has the gap modulation $\delta\Delta>0.03$ meV.

\section{Numerical Results for Landau Levels Spectrum and Conductivity Oscillation}
\begin{figure}
\centering
\includegraphics[width=7in]{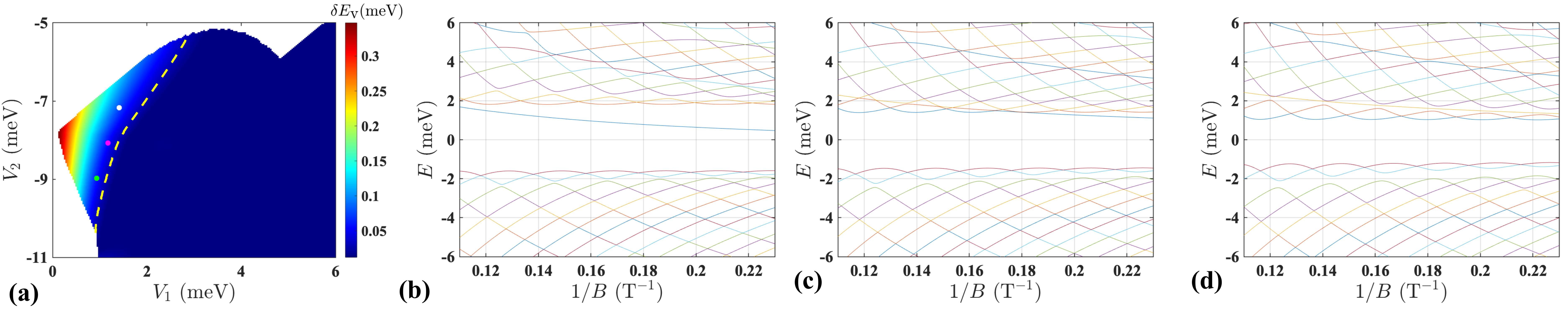}
\caption{(a) The energy difference $\delta E_{\textrm{v}}$ as a function of the coupling potential $\left(V_1, V_2\right)$. The yellow dashed line is the contour $\delta E_{\textrm{v}}=0.3$ meV. The Landau levels spectrum for the white, magenta, and green points are plotted in (b), (c), and (d) respectively. The chemical potentials are also taken to make the oscillation frequency be $f=48.6$ T.}\label{figS3}
\end{figure}

\begin{figure}
\centering
\includegraphics[width=5.8in]{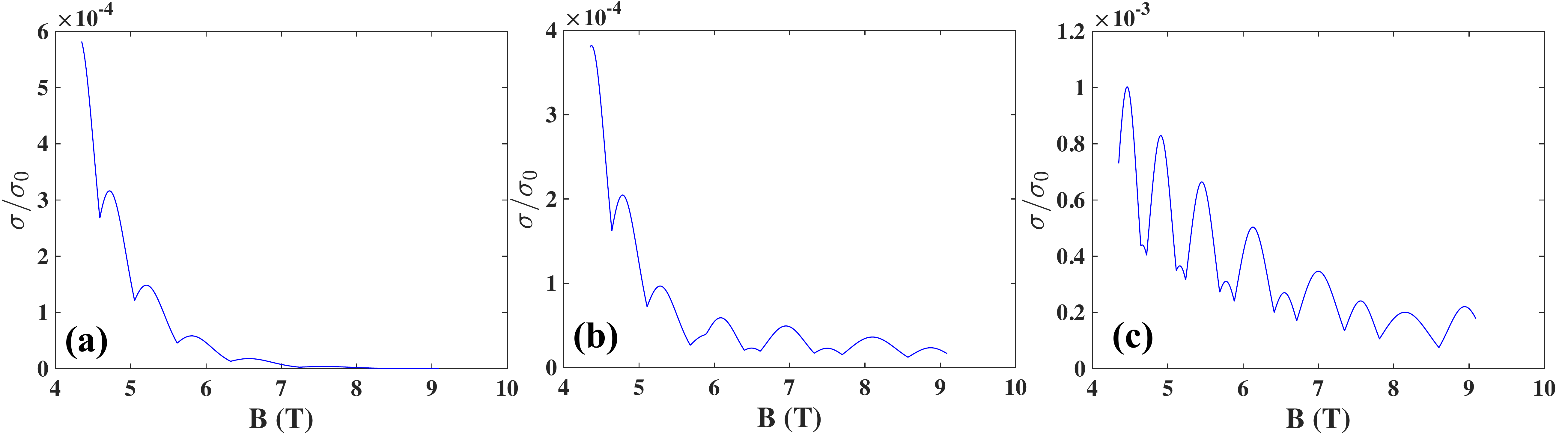}
\caption{The thermally activated conductivity at $T=2$ K. The plots in (a), (b) and (c) are calculated from the Landau levels spectrum in Fig. \ref{figS3} (b), (c) and (d) respectively.}\label{figS4}
\end{figure}

The Landau levels spectrum and the resulting conductivity oscillation have been calculated in a range of parameters. For the relative anisotropy $\gamma=1.5$ case considered in the main text, phase points near the boundary of the regime $\delta E_{\textrm{v}}>0.03$ meV are considered. The Landau levels spectrum all show reasonable gap modulation and the thermally activated conductivity clearly shows oscillation behavior, as seen in Fig. \ref{figS3} and Fig. \ref{figS4}.

\begin{figure}
\centering
\includegraphics[width=5.8in]{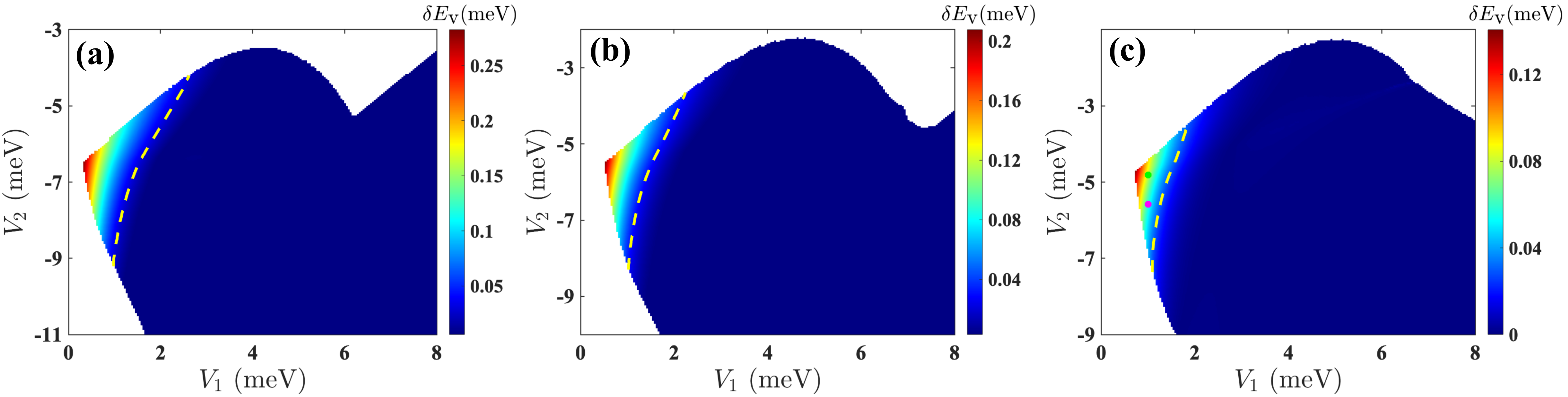}
\caption{The energy difference $\delta E_{\textrm{v}}$ in terms of the coupling potentials $\left(V_1, V_2\right)$. The plots in (a), (b), (c). are for the relative anisotropy being $\gamma=2, 2.5, 3$ respectively. The yellow dashed line is the contour of $\delta E_{\textrm{v}}=0.3$ meV. The contour line gives the estimated regime that has reasonable gap modulation. In the calculation the chemical potentials are fixed so that the oscillation frequency takes $f=48.6$ T.}\label{figS5}
\end{figure}

\begin{figure}
\centering
\includegraphics[width=7in]{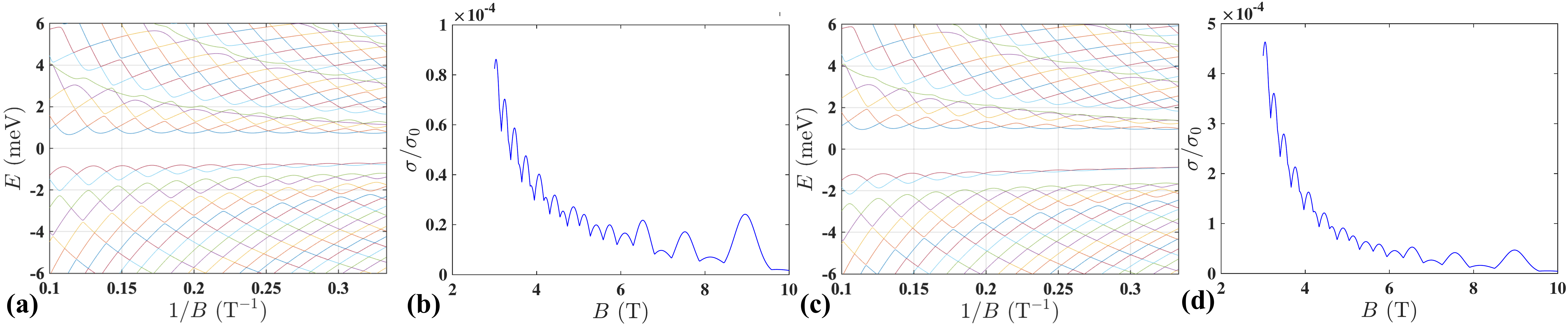}
\caption{The Landau levels spectrum and thermally activated conductivity corresponding to the phase points in Fig. \ref{figS5} (c). The landau levels in (a) and the conductivity oscillation in (b) are from the green point, while those in (c) and (d) are from the magenta point. The temperature is taken to be $T=2$ K for the thermally activated conductivity calculation.}\label{figS6}
\end{figure}

Besides the case of relative anisotropy $\gamma=1.5$, we also give the estimated regime that has reasonable gap modulation for $\gamma=2, 2.5, 3$ in Fig. \ref{figS5}. One can see that the regime of reasonable gap modulation shrinks as the relative anisotropy $\gamma$ increases. In larger relative anisotropy case, the parameter space that has well nested Fermi pockets prior to $V_1$ hybridization shrinks. Then each Landau level from the hole band couples to many other Landau levels in the conduction band, as is indicated by the coupling matrix in Eq. \ref{Coupling_V1}. As a result, the energy oscillation in the resulting top two valence Landau levels gets gradually smoothed. In Fig. \ref{figS6}, Landau levels spectrum and the resulting conductivity oscillation are plotted for two phase points in the phase diagram of $\gamma=3$ case.  The Landau levels spectrum have the gap modulation at the order of 0.1 meV and the resulting conductivity shows observable oscillation.

\begin{figure}
\centering
\includegraphics[width=5.8in]{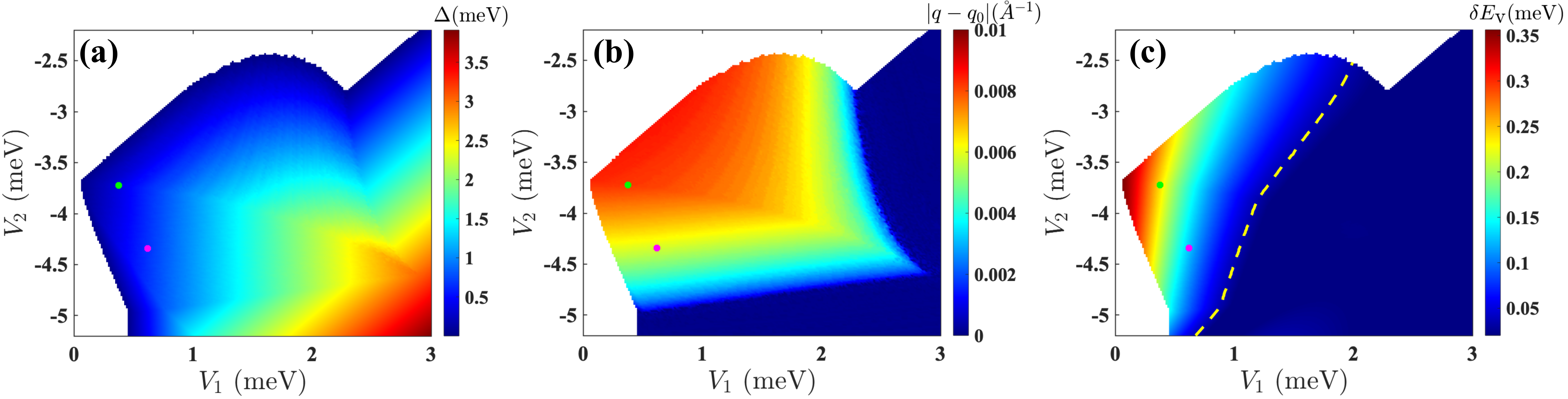}
\caption{The gap size $\Delta$, shift momentum, and energy difference $\delta E_{\textrm{v}}$ in terms of $\left(V_1, V_2\right)$. The chemical potentials are $\mu_{\textrm{c}}=4.6$ meV, $\mu_{\textrm{v}}=3.9$ meV, which makes the oscillation frequency be $f=23$ T. The relative anisotropy is considered to be $\gamma=1.5$.}\label{figS7}
\end{figure}

\begin{figure}
\centering
\includegraphics[width=7in]{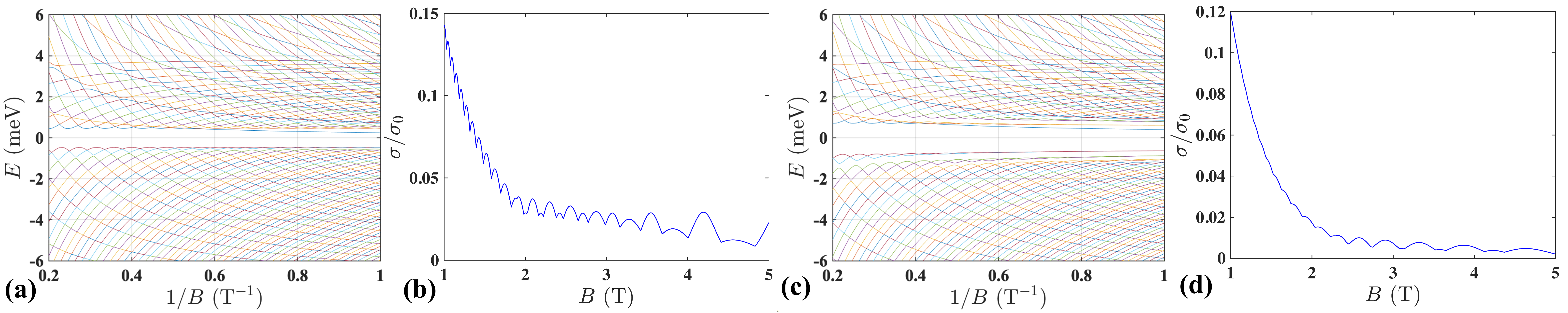}
\caption{The Landau levels spectrum and thermally activated conductivity corresponding to the phase points in Fig. \ref{figS7} (c). The landau levels in (a) and the conductivity oscillation in (b) are from the green point, while those in (c) and (d) are from the magenta point. The temperature is taken to be $T=2$ K for the thermally activated conductivity calculation.}\label{figS8}
\end{figure}

In addition to the oscillation frequency $f=48.6$ T case, which corresponds to the device 1 in the experiment, we also calculate the phase diagram for frequency $f=23$ T case that is related to the device 2 in experiment. We consider the relative anisotropy being $\gamma=1.5$ and the phase diagram is obtained in Fig. \ref{figS7}. In Fig. \ref{figS7}, two phase points inside the reasonable gap modulation regime are selected to calculate the Landau levels spectrum and conductivity. The hybridization gap boundary in Fig. \ref{figS8} (a) and (c) shows periodic oscillation in $1/B$ and it results in the conductivity oscillation in Fig. \ref{figS8} (b) and (d).

\end{document}